\documentclass[12pt]{article}
\usepackage{amsmath} 
\usepackage{multirow} 
\usepackage{amsfonts}
\usepackage{amssymb,graphics,psfrag}
\usepackage{array,epsfig,multirow,stmaryrd,graphicx}
\usepackage{comment}
\def\hybrid{\topmargin -20pt    \oddsidemargin 0pt
        \headheight 0pt \headsep 0pt
        \textwidth 6.25in       
        \textheight 9.03 in       
        \marginparwidth .875in
        \parskip 5pt plus 1pt 
          \jot = 1.5ex
   }
\hybrid
\numberwithin{equation}{section}
\numberwithin{table}{section}

\newcommand{\beq}{\begin{equation}}
\newcommand{\eeq}{\end{equation}}
\newcommand{\be}{\begin{equation}}
\newcommand{\ee}{\end{equation}}
\newcommand{\bea}{\begin{eqnarray}}
\newcommand{\eea}{\end{eqnarray}}   
\newcommand{\ben}{\begin{eqnarray*}}
\newcommand{\een}{\end{eqnarray*}}                  
\newcommand{\ba}{\begin{aligned}}
\newcommand{\ea}{\end{aligned}}
\newcommand{\bt}{\begin{tabular}}
\newcommand{\et}{\end{tabular}}
\newcommand{\bc}{\begin{center}}
\newcommand{\ec}{\end{center}}

\newcommand{\cO}{\mathcal{O}}

\newcommand{\cC}{\mathcal{C}}

\newcommand{\cL}{\mathcal{L}}

\newcommand{\cK}{\mathcal{K}}
\newcommand{\cN}{\mathcal{N}}

\newcommand{\cF}{\mathcal{F}}

\newcommand{\cR}{\mathcal{R}}

\newcommand{\cV}{\mathcal{V}}

\newcommand{\cM}{\mathcal M}

\newcommand{\I}{\text{Im}}
\newcommand{\R}{\text{Re}}

\newcommand{\bj}{{\bar\jmath}}

\newcommand{\Kh}{{\hat{K}}}
\newcommand{\Lh}{{\hat{L}}}
\newcommand{\Ah}{{\hat{A}}}
\newcommand{\Bh}{{\hat{B}}}
\newcommand{\Ch}{{\hat{C}}}
\newcommand{\Dh}{{\hat{D}}}

\newcommand{\ah}{{\hat{a}}}
\newcommand{\bh}{{\hat{b}}}

\newcommand{\bbZ}{\mathbb{Z}}

\newcommand{\bbP}{\mathbb{P}}

\newcommand{\nn}{\nonumber}

\newcommand{\cref}{{\bf [check ref]}}

\newcommand{\simga}{\sigma}

\psfrag{n1}{$\nu_1$}
\psfrag{n2}{$\nu_2$}
\psfrag{n1'}{$\nu_1'$}
\psfrag{n2'}{$\nu_2'$}
\psfrag{n9}{$\nu_9$}
\psfrag{n10'}{$\nu_{10}'$}
\psfrag{t1}{$\tilde{\nu}_1$}
\psfrag{t2}{$\tilde{\nu}_2$}
\psfrag{t9}{$\tilde{\nu}_9$}
\psfrag{t1'}{$\tilde{\nu}_1'$}
\psfrag{t2'}{$\tilde{\nu}_2'$}
\psfrag{t10'}{$\tilde{\nu}_{10}'$}

\def\blfootnote{\xdef\@thefnmark{}\@footnotetext} 
\long\def\symbolfootnote[#1]#2{\begingroup%
\def\thefootnote{\fnsymbol{footnote}}\footnote[#1]{#2}\endgroup}

\begin{document}

\baselineskip=18pt

\begin{titlepage}
\begin{flushright}
\parbox[t]{1.8in}{
BONN-TH-2007-11\\
MAD-TH-07-11\\
0710.3883\ [hep-th]}
\end{flushright}

\begin{center}

\vspace*{ 1.2cm}

{\large \bf  Axion Inflation in Type II String Theory}

\vskip 1.2cm

\begin{center}
 \bf{Thomas W.~Grimm} \footnote{grimm@physics.wisc.edu}
\end{center}
\vskip .2cm

{\em Physikalisches Institut der Universit\"at Bonn, \\[.1cm]
Nussallee 12, 53115 Bonn, Germany}
\vspace*{.1cm}

and
\vspace*{.1cm}

{\em Department of Physics, University of Wisconsin, \\[.1cm]
        Madison, WI 53706, USA
        }

 \vspace*{1cm}

\end{center}

\vskip 0.2cm

\begin{center} {\bf ABSTRACT } \end{center}

Inflationary models driven by a large number of axion fields are discussed
in the context of type IIB compactifications with $\cN=1$ supersymmetry. 
The inflatons arise as the scalar modes of the R-R two-forms evaluated 
on vanishing two-cycles in the compact geometry. The vanishing 
cycles are resolved by small two-volumes or NS-NS B-fields 
which sit together with  the inflatons in the same supermultiplets. 
String world-sheets wrapping the vanishing cycles 
correct the metric of the R-R inflatons. They can help to
generate kinetic terms close to the Planck scale and a mass hierarchy between the 
axions and their non-axionic partners during inflation.
At small string coupling, D-brane corrections are subleading 
in the metric of the R-R inflatons.  However, an axion potential can be 
generated by D1 instantons or gaugino condensates on D5 branes. 
Models with sufficiently large number of axions admit 
regions of chaotic inflation which can stretch over the whole axion
field range for potentials from gaugino condensates.
These models could allow for a possibly detectable 
amount of gravitational waves with tensor to scalar ratio as high 
as $r<0.14$.

\end{titlepage}


\section{Introduction}

Current observational cosmology allows us to test fundamental physics 
with a continuously improving precision. To evaluate and understand 
these data, theoretical models about the evolution of our universe are
crucial. One promising paradigm receiving growing experimental 
support is cosmological inflation \cite{original_inflation}. Inflation
postulates a period of exponential expansion of the universe driven 
by scalar fields slowly rolling in an almost flat potential. 
This enormous growth 
stretches quantum fluctuations present in the early universe 
to currently observable astrophysical scales. The imprints 
of such a process can be found, for example, in the cosmic
microwave background and the large scale structure of the 
universe \cite{WMAP,Tegmark, Sanchez}.

In recent years much effort has focused on the realization 
of inflation within string theory \cite{Cosmo_rev,Hertzberg:2007ke}.
In string theory there are
potentially many scalar fields which could drive inflation and hence 
different opportunities to model inflation. Specific scenarios include 
realizations of K\"ahler moduli inflation~\cite{Banks:1995dp,Conlon:2005jm}, 
racetrack models \cite{BlancoPillado:2006he} and the most intensively 
studied possibility of D-brane inflation \cite{Dvali:1998pa,Cosmo_rev}. 
However, as of today, it remains challenging to establish explicit scenarios
in a controlled compactification without employing extreme fine-tuning 
to obtain a sufficiently flat 
potential \cite{BDKMcAS}. Having found a realization of inflation reproducing the current 
cosmological observables, it is important to establish which models 
can incorporate possible future observations. For example, as argued in refs.~\cite{BMcA}, many 
string scenarios do not allow for a high ratio $r$ of gravitational waves 
produced in the early universe. The current experimental bound on  
gravitational waves is $r<0.3$ \cite{Tegmark}, but future experiments, including 
Planck, BICEP and Spider \cite{Efstathiou:2006ak},
might allow the observation of $r$
with a precision down to $r >0.01$. It is thus desirable to 
study string embeddings of inflation which can incorporate an $r$ 
observable in these experiments. Recent attempts to do that can be  
found, for example, in refs.~\cite{Olsson:2007he,Krause:2007jr,Becker:2007ui,Kobayashi:2007hm, KSS}.

A possible scenario able to incorporate primordial gravitational waves 
was suggested by Dimopoulos, Kachru, McGreevy 
and Wacker \cite{DKMW}.  The authors argue for an 
embedding of multi-field inflation with a large number $N$ 
of axion fields. Such models use an assistance 
effect studied in refs.~\cite{LMS,KO} ensuring that the small fraction
$1/N$ controls the flatness of the potential.
Indeed, generic compactifications of type II string theory on a 
six-dimensional manifold can admit  $10^4$, or more,
axions from the NS-NS B-field and the R-R form fields.
An appropriate subset of $N$ such 
axions were proposed to  drive inflation in ref.~\cite{DKMW} 
and the authors termed these scenarios 
$N$-flation.
For a sufficiently large $N$ the scenarios can be 
interpreted as a realization of natural inflation \cite{Freese:1990rb,Kim:2004rp}.
If the inflatons can produce the desired amount of e-foldings
already in the quadratic regime of the potential 
the models admit chaotic inflation \cite{Linde:1983gd}.
This implies that $N$-flation can yield a possibly 
observable signature of gravitational waves with 
$r<0.14$ and thus distinguishes it from most other string realizations of inflation.

The aim of this paper is to study a specific realization of 
$N$-flation in type IIB string theory. The inflating 
axions correspond to the zero-modes of the R-R forms 
in the compactification to four space-time dimensions. 
In compactifications preserving $\cN=2$ supersymmetry 
the R-R axions sit  in the same supermultiplets as the K\"ahler structure 
moduli parametrizing the volumes of two-dimensional 
cycles in the internal space. Manifolds with a large 
number of  non-trivial two-cycles are thus candidate backgrounds
for $N$-flation. As will be shown, the density perturbations and 
slow roll parameters depend on the volume of the compact 
space and it has to be ensured that they do 
not become large with increasing $N$.
We will thus argue that explicit examples always involve compact
manifolds which admit many very small or vanishing cycles. 

In the presence of small cycles stringy effects become important and significantly alter 
the structure of the four-dimensional effective theory.  
In order to analyze these contributions we will concentrate on
axions arising from the R-R two-form evaluated on  vanishing two-cycles 
of the compact geometry. The standard example of a vanishing 
two-cycle is the resolved conifold \cite{Candelas:1989js}. In this case a conical singularity is 
resolved by a two-sphere supported by a geometric volume or an
NS-NS B-field. If this $S^2$ becomes smaller than the 
square string length, world-sheets will start  to wrap and significantly
contribute to the metric of the R-R axions. Fortunately, in $\cN=2$ 
compactifications these corrections can be computed for the conifold 
and many other Calabi-Yau geometries \cite{Hori:2003ic,Hosono:1993qy} allowing the evaluation of
the metric of the axions. We will illustrate this general fact on a toy model 
with $N$ conifold singularities. For such examples it can be shown that 
the kinetic terms of the axions can be close to the Planck scale which is 
crucial to obtain inflation \cite{BDFG}. 

In addition to the fundamental strings also D-branes can become
relevant in geometries with vanishing cycles \cite{Strominger:1995cz}. In particular, 
D1 instantons can wrap the small cycles and correct the metric 
of the R-R axions. Such contributions appear with the 
exponential of the D1 instanton action which depends 
on the R-R two-form axions themselves. In contrast to the string 
world-sheet action the D-instanton action also contains a factor of the inverse 
string coupling. This implies that
for small string coupling and finite volume or B-field of
the vanishing cycles the D1 instanton corrections are subleading 
in the axion metric. However, correction due to D-branes will be the 
leading contributions in the scalar potential and can 
induce the desired potential for the R-R axions.

To study the scalar potential for the axions and non-axionic 
moduli we will focus on $\cN=1$ orientifold compactifications of type IIB string 
theory. Such compactifications have been studied intensively in the last years \cite{review_flux}. 
It was shown that $\cN=1$ potentials can 
be induced by background fluxes, D-brane instantons or 
gaugino condensates on space-time filling D-branes.
The desired axion potentials can arise through non-trivial 
superpotentials from either of the three sources \cite{review_flux}. 
We will briefly discuss their properties in various orientifold 
compactifications: D1 superpotentials in type I~\cite{Witten2}, 
D1 dependences through the determinants in the D3 instanton 
superpotentials \cite{Witten3,Ganor,TG},
and gaugino condensates on space-time filling D5 branes 
\cite{review_flux,Vafa:2000wi,Cachazo:2001jy,Dijkgraaf:2002dh}.
Remarkably, explicit computations of the non-perturbative superpotentials
can often be performed in a dual flux picture where geometry dictates
the form of the corrections
 \cite{Vafa:2000wi,Cachazo:2001jy,Dijkgraaf:2002dh,Heckman:2007ub,ABK}.

In the final part of this work we will study the effective theory 
of orientifold compactifications with O3 and O7 planes 
in more detail.  Using earlier results \cite{GL1,GL2} we 
argue that the $\cN=2$ world-sheet corrections are inherited 
by the $\cN=1$ orientifold theory.  In particular, they 
correct the $\cN=1$ K\"ahler potential and complex coordinates
in a calculable way. This general fact can be applied to a simplistic 
compact toy model with $2N$ conifolds pairwise identified under the orientifold 
projection. Including a flux and D-instanton superpotential, we make first 
steps in establishing an effective theory with a large number of 
light axions 
and all other moduli stabilized. 
An explicit numerical evaluation indicates the presence of 
a non-supersymmetric axion valley \cite{Kallosh}. 
This ensures the desired mass hierarchy between 
the axions and their non-axionic partners.  

The paper is organized as follows. In section \ref{rev_N} we review 
the $N$-flation scenario of \cite{DKMW} and 
discuss some of its cosmological implications.
We recall that the kinetic terms of the axions, set by the axion decay 
constants, have to be large in order to obtain inflation. A discussion of 
axion decay constants in Type IIB string theory is presented in section~\ref{axion_decay}.
The four-dimensional effective Lagrangian and the general 
form of the axion decay constants are studied in section \ref{four-dim_Lagr} supplemented 
with appendix \ref{gen_axion_const}.
In section \ref{large_vol} we argue that in compactifications with all cycles larger than string length,
the axion decay constants typically become very small with an increasing number 
of axions. This implies that only compactification manifolds with small or vanishing cycles are
candidate backgrounds to obtain $N$-flation. The quantum corrected 
axion decay constants for the resolved conifold and geometries with 
$N$ resolved singularities are discussed in sections \ref{res_cone} and \ref{multiax}. 
Non-perturbative D-brane effects can induce 
a scalar potential through a non-vanishing superpotential as discussed in section \ref{axion_potential}. In section \ref{D1_corrections}
it is shown that such a superpotential can arise from D1 instanton 
corrections, while section~\ref{D5_gaugino} discusses superpotentials originating from gaugino 
condensates on D5 branes.
In the final part of this work, section \ref{N=1_ori}, the embedding of $N$-flation into a concrete 
$\cN=1$ orientifold compactification is addressed. The general 
form of the $\cN=1$ data including the inherited $\cN=2$ 
perturbative and non-perturbative string world-sheet corrections 
is presented in section~\ref{four-dim_N=1} and appendix~\ref{N=1_recall}.
In section~\ref{N_conifolds}, a toy model with $N$ conifold pairs is
used to illustrate that the string world-sheet corrections in the K\"ahler potential and the presence
of a non-perturbative superpotential can ensure an effective theory with light axions.
This indicates the possibility of $N$-flation in string theory.

\section{Review of axion $N$-flation \label{rev_N}}

In this section we will review some basics about N-flation driven by a large 
number of axion fields following \cite{DKMW, EMcA}. 
The basic idea is that by increasing the set of inflaton fields an assistance 
effect can help to ensure that the slow roll conditions are met \cite{LMS,KO}. More precisely, one considers 
set-ups with inflaton fields $c^a,\ a=1\ldots N$, where each field feels the downward force of its own potential 
but is slowed down by the collective frictional force of all fields. 
To make this more explicit, let us consider a set of inflaton fields $c^a$ with Lagrangian 
\beq \label{simp_act}
  \cL = \tfrac{1}{2} f^2_{ab}\, \partial_\mu c^a \partial^\mu c^b - V + \ldots  \ .
\eeq 
To employ the assistance effect we will specify the set-up further. 
We consider a scenario, where, at least approximately, the metric $f^2_{ab}$ is independent 
of $c^a$ and can be diagonalized to have a diagonal $f^2_{aa} =f^2_a$. In order to obtain canonically normalized
kinetic terms in \eqref{simp_act} we introduce the fields $\theta^a$
\beq \label{def-theta}
    c^a= \theta^a/f_a   \ , \qquad a=1\ldots N\ .
\eeq
Moreover, we also constrain the potential for the fields $\theta^a$.
We assume that in an effective description
the potential $V \approx V_{\rm eff}$ is given by 
\beq \label{split_potential}
    V_{\rm eff}(\theta^a) = \sum_{a} V_a(\theta^a)\ ,
\eeq  
where each term $V_a(\theta^a)$ only depends on the 
$a$th inflation field $\theta^a$. For a time-dependent evolution in a Friedmann-Robertson-Walker universe
the equations of motion for the fields $\theta^a$ take the
simple form 
\beq \label{eom}
  \ddot \theta^a + 3 H \dot \theta^a + \partial_a V = 0 \ ,\qquad H^2 = \frac{1}{3 M^2_P}\big( \tfrac{1}{2}(\dot \theta^a)^2 + V \big) \ .
\eeq
The assistance effect is now apparent. The Hubble friction contains the whole potential 
of all fields $\theta^a$ while the downward force $\partial_a V$ only yields a 
non-vanishing contribution from the $a$th potential term in \eqref{split_potential}.
For a large number of fields this assistance can help to ensure slow roll for the
inflaton fields $\theta^a$.

\begin{figure}[!ht]
\leavevmode
\begin{center}
\includegraphics[height=4.1cm]{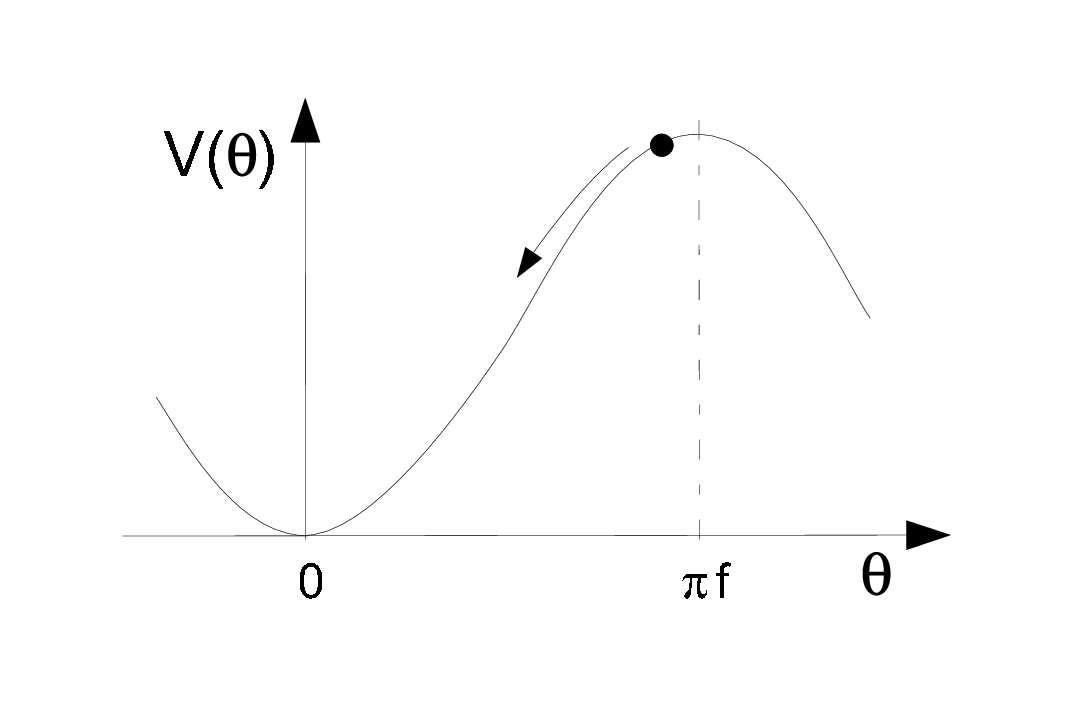} 
\end{center}
\vspace*{-.5cm}
\caption{\small Potential for one axion field $\theta$ with $\mu=1$. }
\label{Potential}
\end{figure}

In the following we will be more concrete and study an explicit potential in more detail. 
Our aim is to identify $c^a=\theta^a/f_a$ with 
axion fields in a string compactification. The axions $c^a$ are periodic with 
period $2\pi$ such that the accessible field range are the intervals 
\beq \label{intervals}
   -\pi\ <\ c^a \ \le\ \pi \ , \qquad \quad -f_a \pi \ <\ \theta^a \  \le\ f_a \pi\ .
\eeq
A potential term for the 
axions arises only through non-perturbative corrections  to the four-dimensional 
effective theory as will be discussed in section \ref{axion_potential}. This implies
that the approximate effective potential \eqref{split_potential} for the axion fields is of the form \footnote{%
The potential generally also contains cross 
coupling terms of the form $\cos \big[\mu^a\, \theta^a/f_a -\mu^b\, \theta^b/f_b \big]$ as discussed in refs.~\cite{DKMW,EMcA}.
These will be omitted in this section.}
\beq \label{sum-pot}
  V_{\rm eff}(\theta^a) = C + \sum_{a=1}^N \Lambda^4_a\big(1- \cos \big[\mu^a\, \theta^a/f_a \big] \big)\ . 
\eeq
Let us introduce the different variables appearing in $V_{\rm eff}$. The constants $f_a$ 
arise through the redefinition \eqref{def-theta} ensuring that $\theta^a$ have canonically normalized 
kinetic terms. $f_a$ are the axion decay constants appearing in the metric \eqref{simp_act}.
Non-trivial $\mu^a$ might already arise in 
the potential for $c^a$ and thus do not appear due to the rescaling of the fields. 
In section \ref{D5_gaugino} it will be analyzed how non-trivial $\mu^a$ arise in string 
theory. The constants $\Lambda_a$ set the scale of inflation and are typically determined by 
the vevs of other fields in the full string compactification. 
The constant $C$ is the value of the effective potential 
at the minimum where all $\theta^a = 0$. This cosmological constant $C$ is, in accordance with current observations, very small 
and can be safely approximated to be zero for the following analysis. Figure \ref{Potential} shows one of the periodic potentials
of the sum \eqref{sum-pot}.

Let us now discuss inflation driven by the scalar fields $\theta^a$ in the potential \eqref{sum-pot}.
To obtain slow roll inflation we need to satisfy the standard slow roll conditions \cite{Linde:2005ht,Lyth:1998xn}. We will first 
introduce the slow roll parameters for a separable potential $V$ as in \eqref{split_potential}.
In this case the 
slow roll parameters are given by 
\beq \label{slow_roll}
   \epsilon = \frac{M_P^2}{2} \sum_a\ \left( \frac{V_{,a}}{V}\right)^2\ ,\qquad
    \eta =    M_P^2\ \mathop{\text{min}}_{a}\left( \frac{V_{,aa}}{V}\right)\ .
\eeq
where $V_{,a}\equiv \partial_{\theta^a} V$ and $V_{,aa}= \partial^2_{\theta^a} V$.
The slow roll conditions read $\epsilon < 1$ and $|\eta| <1$ and define a multi-dimensional 
subspace in the fields $\theta^a$ where inflation takes place. In this inflationary region of the field-space
the Hubble friction \eqref{eom} is well approximated by the potential 
$H = V / (3 M^2_P)$ and the physical observables 
can be defined as a function of the potential $V$ and its derivatives.

The number of e-foldings and the magnitude of scalar density perturbations 
during the slow roll epoch is given by 
\beq \label{Nr_efolds}
   N_e = -\sum_a \int_{\theta_{\rm in}}^{\theta_{\rm fin}} \frac{V_a}{V_{,a}} \frac{d\theta^a}{M^2_P}\ ,\qquad \qquad \quad
   \left( \frac{\delta \rho}{\rho}\right)^2 = \frac{V}{75 \pi^2 M_P^6} 
   \sum_a \left(\frac{V_a}{V_{,a}}\right)^2 \ .
\eeq
$N_e$ parametrizes the exponential growth of the universe during inflation. It has to 
be sufficiently large, $N_e \gtrsim 50$, to ensure that different parts of the early universe 
have be in causal contact.  
The density perturbations seen in the CMB arise from the time $50-60$ e-foldings before the end of inflation. For simplicity, we will take $\theta_{\rm in}^a \approx \theta_{\rm in}$ to be the starting point of $55$ e-foldings 
of inflation and evaluate all cosmological observables around this point in field-space.
In accordance with current observations, we demand
\beq \label{obs_Ne}
     N_e \approx 55\ ,\qquad \qquad   (\delta \rho/\rho )_{\theta_{\rm in}} \approx 2 \times 10^{-5}\ .
\eeq
Similarly, the other cosmological observables can be defined for multi-field 
scenarios. We will not review all the details here, but rather 
refer to the literature for a more exhaustive discussion \cite{Lyth:1998xn}. 
There is, however, one more observable which we like to introduce.
We denote by $r= P_g/P_\cR $ the relative magnitude of gravity waves $P_g$ to 
density perturbations $P_{\cR}$. Recalling that $P_g$ and $P_{\cR}$
are given by
\beq
   P_{g} = \frac{2}{3 \pi^2} \frac{V}{M_{P}^4}\ , \qquad \qquad \quad P_{\cR} = \frac{25}{4} \left(\frac{\delta \rho}{\rho} \right)^2\ ,
\eeq
the ratio $r$ is seen to be
\beq \label{def_r}
   r = 8 M_P^2 / \sum_a  \left(\frac{V_a}{V_{,a}}\right)^2 \ .
\eeq
The current bounds on $r$ demand that 
\beq \label{obs_bound_r}
   (r)_{\theta_{\rm in}} < 0.3 \ ,
\eeq 
where $r$ is evaluated close to $55$ e-foldings before the end of inflation.
However, future experiments might show 
that $r$ is not much below this bound. As we will discuss momentarily, due to the assistance 
effects of the many axions, $r$ can be close to this bound in models of axion N-flation.
This provides one of the major motivations to study these scenarios, since typically 
$r$ in string motivated models such as D-brane inflation \cite{Cosmo_rev} takes a 
rather small value \cite{BMcA}.

In the remainder of this section we will evaluate the slow roll parameters $\epsilon,\ \eta$
as well as the number of e-foldings $N_e$ and the ratio $r$ for the potential \eqref{sum-pot}. Clearly, using 
\eqref{slow_roll}, \eqref{Nr_efolds} and \eqref{def_r} this is straightforward and can be done in full generality. 
However, to simplify the analysis and to make the results more transparent we will 
set all axion decay constants $f_a$ and the coefficients $\mu^a$ to be approximately 
of the same size, and set $\mu^a/f_a \approx A$. We also assume 
that $\Lambda_a \approx \Lambda$. 
Now we evaluate the slow roll parameters \eqref{slow_roll} for $\theta^a \approx \theta$ by
using the effective potential \eqref{sum-pot}. Explicitly, we find 
\beq \label{eps_eta_N}
     \epsilon  =  \frac{M_P^2}{2}  \frac{A^2\, \sin^2 (A \theta)}{N \big(1- \cos(A \theta) \big)^2} \ , \qquad \qquad 
     \eta  = M_P^2  \frac{A^2\, \cos(A \theta)}{N \big(1- \cos(A \theta) \big)} \ .
\eeq
An immediate conclusion is that $\epsilon$ as well as $\eta$ are independent of the scale $\Lambda$, but 
crucially depend on the constants $A$ and the number of axions $N$.
The number of e-foldings and amount of scalar density perturbations \eqref{Nr_efolds} are 
given by
\beq \label{N_e_N}
   N_e = -\frac{2 N}{M_P^2\, A^2} \text{log}\left[\frac{\cos(A \theta_{\rm in}/2)}{\cos(A \theta_{\rm fin}/2)} \right] \ ,\qquad \quad
   \frac{\delta \rho}{\rho} = \frac{2 \Lambda^2 N}{M_P^3} \frac{\big(1- \cos(A \theta_{\rm in}) \big)^{3/2}}{A \sin (A \theta_{\rm in})}\ ,
\eeq
where $\theta_{\rm in}$ and $\theta_{\rm fin}$ is the starting and endpoint of inflation. 
Finally, we will derive the 
relative magnitude of gravity wave to 
density perturbations $r$. Inserting \eqref{eps_eta_N} into \eqref{def_r} one finds with the above simplifications
\beq \label{ev_r}
    r=16 \epsilon= 8 M_P^2  \frac{A^2\, \sin^2 (A \theta_{\rm in})}{N \big(1- \cos(A \theta_{\rm in}) \big)^2} \ .
\eeq
From this expression we infer that $r$ will be small if $\theta_{\rm in}$ is 
close to the maximum of the potential. However,
for sufficiently large $N$ and small $A$ horizon crossing can take place away from the maximum. 
In such cases, $r$ can be close to the current observational bound \eqref{obs_bound_r}. This is particularly easy to 
see in the quadratic regime of the potential which we will study in the finial part of this section.

For a subset of values $N$ and $f_a,\ \mu^a$ a quadratic approximation of the potential \eqref{sum-pot} suffices for slow roll inflation. 
More precisely, we will concentrate on the regime 
\beq  \label{quad_reg}
    \mu^a \theta^a/ f_a \approx A \theta^a < 1\ ,\qquad \qquad V_{\rm eff}(\theta^a) \approx m^2 \sum_a (\theta^a)^2\ ,
 \eeq
where $m^2 = \Lambda^4 A^2$ are the masses of the axions $\theta^a$ which are assumed to be similar in magnitude.
 In this regime we can expand the trigonometric functions in \eqref{N_e_N} and \eqref{ev_r} to lowest order in the 
 $\theta$'s. One obtains a model of chaotic inflation for which the assistance effect of the $N$ inflatons 
 allows effectively super-planckian vev's. To qualitatively discuss this regime we again assume $\theta^a\approx\theta=\alpha M_P$
 and include appropriate factors of $N$ in our analysis. In this limit \eqref{eps_eta_N} and \eqref{N_e_N} simplify to
 \beq \label{eps_eta_quad}
  \epsilon \approx \frac{2}{N \alpha^2}\ ,\qquad   \qquad 
  \eta \approx  \frac{2}{N \alpha^2}\ ,
 \eeq
 and  
 \beq \label{Nrhor_quad}
 N_e \approx \frac{N \alpha^2_{\rm in}}{4}\ , \qquad \qquad 
   \frac{\delta \rho}{\rho}\approx\frac{m\, N \alpha^2_{\rm in}}{M_P} \ ,\qquad  \qquad 
   r \approx \frac{32}{N \alpha^2_{\rm in}} \ .    
 \eeq
 Using \eqref{obs_Ne} we conclude that $N \alpha^2_{\rm in}\approx 220$ and $m/M_P \approx 10^{-7}$. 
 Evaluating $r$ using these results gives $r\approx 0.14$ which is not significantly below the 
 current bound \eqref{obs_bound_r}. In case we are able to realize axion inflation in a controlled string 
 compactification it might be the ratio $r$ which crucially distinguishes it from other scenarios 
 such as the inflation of a small number of D-branes \cite{BMcA}. 
 
Let us 
 note that in order to obtain slow roll  
 in the quadratic regime \eqref{quad_reg} we have to ensure that $N$ is very large and 
 $\mu_a/f_a$ is small. In particular, it has to be guaranteed that 
 the axions generate a sufficient number of e-foldings, $N_e \approx 55$, during 
 the slow roll phase in the regime \eqref{quad_reg}.
However, for $\mu_a=1$ a significant
part of the e-foldings are generated close to the maximum of the cosine in \eqref{sum-pot},
where the quadratic approximation is no longer valid. 
Including the constants $\mu_a$ in the discussion is more subtle,
as we will discuss in more detail in section \ref{D5_gaugino}. The reason is that
a change in $\mu_a$ does not alter the field range \eqref{intervals} of 
the canonically normalized axions. Nevertheless, the quadratic 
approximation \eqref{quad_reg} becomes valid for more values of $\theta^a$ 
if $\mu_a \le 1$ is small. Eventually, small values of $\mu_a$ can push 
the maximum out of the field range of $\theta^a$ and even ensure that a 
quadratic approximation is valid for all values of $\theta^a$ in \eqref{intervals}.
Both in the quadratic regime as well as for the full potential, 
we observe that for large $N$ and $f_a$  the slow roll condition $\epsilon < 1,\ \eta < 1$ is easier to 
satisfy for many values of $\theta^a$. We thus have to ensure that sufficiently large 
values for $N$ and $f_a$ are accessible in a string embedding.

In the next section we will start the discussion of axion inflation in 
type IIB string theory by evaluating the axion decay constants for different
R-R and NS-NS axions. We will recall that in a controlled string compactification 
$f_a \lesssim M_P$ seems unavoidable.  In order to nevertheless obtain a sufficiently 
long period of slow roll inflation, we have to ensure that both $N$ and $f_a$ are 
close to the accessible values.

\section{Axion decay constants in IIB string theory \label{axion_decay}}

In the previous section we reviewed a simple inflationary scenario 
driven by the dynamics of a large number of axions. In order 
to embed such a model into string theory, we 
have to identify an appropriate set of 
axions $\theta^a$ arising in a compactification of string theory 
from ten to four space-time dimensions. As we have seen, the cosmological 
models crucially depend on the number of axions $N$ and 
the value of the axion decay constants $f_a$. 
In order to guarantee an epoch of slow roll inflation 
we have to ensure that both $N$ and $f_a$ take sufficiently large values.
In this section, we discuss the axion decay constants of type IIB compactifications 
on Calabi-Yau manifolds with $\cN=2$ supersymmetry. 
We will see, that in string theory an implementation of axion decay constants 
close to the Planck scale ($f_a \lesssim M_P$), for a large number of axions, 
is in general hard to achieve.  It appears that candidate scenarios can only 
arise away from the large volume limit, where the compact geometry admits
small or vanishing cycles.

\subsection{The four-dimensional axion Lagrangian \label{four-dim_Lagr}}

In this section we will fix our conventions and discuss the four-dimensional effective Lagrangian 
for the NS-NS and R-R axions obtained by compactifying type IIB string theory 
on a Calabi-Yau manifold $Y$. This will allow us to determine 
the axion decay constants for the various types of axions.

Let us begin by summarizing our conventions following \cite{JPbook,SW}.
The ten-dimensional gravitational coupling $\kappa_{10}$ in the Einstein-Hilbert term  
is given by $\kappa_{10}^2 = g_s^2 \ell_s^8/4\pi$, where $\ell_s = 2\pi \sqrt{\alpha'}$ is the 
string length and $g_s$ is the type II string coupling. The string coupling is normalized such that 
S-duality acts as $g_s \rightarrow 1/g_s$. The four-dimensional Planck mass $M_P$ is obtained 
by dimensionally reducing the Einstein-Hilbert term to four 
dimensions. Denoting by $V_Y$ the volume of the internal manifold, $M_{P} = 2 \times 10^{18}\, \text{GeV}$ is given by 
\beq
   M_P^2 = \frac{4 \pi V_Y}{g_s^2 \ell_s^8}\ , \qquad V_Y=R^6\ ,
\eeq
where $R$ denotes the typical radius of the compactification space. It turns out 
to be convenient to introduce the dimensionless volume $\cV= V_Y/\ell_s^6$. The 
ratio of string scale $m_s=1/\ell_s$ to Planck scale is then given by
\beq
   \frac{m_s}{M_{P}}= \frac{g_s}{\sqrt{4 \pi \cV}}\ .
\eeq
This fraction is smaller than unity, since we will consider the volume $\cV$ to be above string scale, i.e.~$\cV>1$,
and $g_s$ to be at weak coupling.  
Following \cite{SW} we will normalize the NS-NS and R-R field strengths to 
have integral periods. This implies that the kinetic terms for the metric as well as the NS-NS and R-R 
fields $B_2,C_p$ are obtained from the ten-dimensional string-frame action of the form 
\beq \label{ten-RR}
  - \frac{2\pi}{\kappa_{10}^2} \int \Big[ \frac{1}{2} e^{-2\phi} R *1+
  \frac{1}{4}\frac{1}{ \ell_s^{4} } e^{-2\phi} H_{3} \wedge * H_{3}   + \frac14 \frac{1}{ \ell_s^{8-2p} }\ F_{p+1} \wedge * F_{p+1}\Big]\ ,
\eeq
where $H_3$ and $F_{p+1}$ are the fieldstrengths of $B_2$ and $C_p$ respectively. 
In the expression \eqref{ten-RR} the field $e^{\phi}$ denotes
the ten-dimensional dilaton with vacuum expectation value $g_s = e^{\langle \phi \rangle}$ being
the string coupling constant.
Finally, in accordance with \eqref{ten-RR}, the coupling of the R-R forms to D$p$-branes take the form $2\pi \int C_p$.
The tension of a D$p$-brane is given by $T_p = 2 \pi/(g_s \ell_s^{p+1})$.

In compactifications of type IIB string theory a large number of  axions 
can only arise from the NS-NS and R-R two-forms $C_2, B_2$ and the R-R four-form $C_4$.
More precisely, taking the internal six-dimensional space to be a Calabi-Yau manifold $Y$,
the axions arise by expanding $B_2, C_2$ and $C_4$ in a basis of harmonic forms. 
Denoting by $\omega_A$ a basis of two-forms in $H^{2}(Y,\bbZ)$ and by $\tilde \omega^A$ its dual basis 
of four-forms in $H^{4}(Y,\bbZ)$ we can expand 
\beq \label{C-expansion}
   B_2 = \tfrac{1}{2 \pi}\, b^A \omega_A \ ,\qquad \quad C_2 = \tfrac{1}{2 \pi}\, c^A \omega_A \ ,\qquad \quad 
   C_4 =\tfrac{1}{2 \pi}\, \rho_A \tilde \omega^A\ ,
\eeq
where the factors of $1/2\pi$ were included to ensure that the axions are $2\pi$ periodic.
We also expand the K\"ahler form $J$ of $Y$ into the above integral basis
\beq \label{J_exp}
  J =  v^A \omega_A= (R^A /\ell_s)^2 \omega_A \ .
\eeq
Here $J$ is normalized to be dimensionless, while the $(R^A)^2$ are the 
dimensionfull volumes of two-cycles. Compactifying type IIB string 
theory on the Calabi-Yau manifold $Y$ yields a four-dimensional 
$\cN=2$ supergravity theory. Indeed, the axion fields in \eqref{C-expansion} 
combine with the K\"ahler structure deformations into $\cN=2$
hypermultiplets with scalars $(v^A,b^A,c^A,\rho_A)$. The effective four-dimensional 
action for these hypermultiplets was derived in \cite{FS}. In the following, 
we will only discuss some of the relevant terms and determine 
the leading axion decay constants.  A more complete study of the action and the axion decay 
constants can be found in  appendix \ref{gen_axion_const}.

As in section \ref{rev_N}, the axion decay constants are determined from the kinetic terms of $b^A, c^A$ and 
$\rho_A$. These are derived by dimensionally reducing the ten-dimensional action \eqref{ten-RR}. 
The resulting four-dimensional Lagrangian for the axions takes the form
\beq \label{C2C4_4d}
  \cL = - \tfrac{M^2_P}{2 \pi} \Big(
 G_{A \bar B} \partial_\mu b^A \partial^\mu b^B   +
  e^{2\phi}\, G_{A \bar B} \partial_\mu c^A \partial^\mu c^B + e^{2\phi}\, \tilde  G^{A \bar B} \partial_\mu \rho_A \partial^\mu \rho_B \Big)\ ,
\eeq
where $G_{A \bar B}$ and $\tilde G^{A \bar B}$ are the moduli space metrics. It is important to note that in the 
expression \eqref{C2C4_4d} we have left out 
additional terms which will not be relevant in the following but are of crucial importance 
for the effective action to have $\cN=2$ supersymmetry (see appendix \ref{gen_axion_const}).  
The leading axion decay constants are 
determined in terms of the metrics $G_{A \bar  B}$ and $\tilde G^{A \bar  B}$ as 
\begin{align} \label{decay}
   B_2:\quad \frac{f^2_{AB}}{M^2_P} = \frac{1}{\pi} \, G_{A \bar B} \ ,\qquad \quad  C_2:\quad \frac{f^2_{AB}}{M^2_P} = \frac{g_s^2}{\pi}\, G_{A \bar  B} \ , \qquad \quad C_4:\quad \frac{f^2_{AB}}{M^2_P} = \frac{g_s^2}{\pi}\, \tilde G^{A \bar  B} \ .
\end{align}
In these expressions we have taken the dilaton to be fixed
to its vacuum value $g_s$.
Clearly, in order to evaluate the typical size of the axion 
decay constants we will need the explicit form of the
moduli space metrics appearing in \eqref{decay}.

The moduli space metrics in \eqref{decay}  are 
functions of the complexified K\"ahler structure deformations
\beq \label{def-t}
   t^A = - b^A + i v^A\ .
\eeq
Due to the underlying $\cN=2$ supersymmetry all dependence of $G_{A\bar B}$
on $t,\bar t$ can be encoded by a holomorphic function $\cF(t)$ known as the 
pre-potential. In general, $\cF$ contains a classical piece cubic  in $t$ as well as perturbative
and non-perturbative string world-sheet corrections of order $\alpha'$.
The metric $G_{A\bar B} = \partial_{t^A} \partial_{\bar t^B} K$ is a K\"ahler metric 
with K\"ahler potential \cite{Candelas:1990pi}
\beq \label{general_K}
  K(t,\bar t) = - \ln \cV\ ,\qquad \cV = 2i(\cF -\bar \cF) - i(\partial_{t^A} \cF + \partial_{\bar t^A}\bar \cF) (t^A - \bar t^A)\ ,
\eeq
where $\cV$ is the quantum corrected volume of the compact Calabi-Yau space $Y$.
Up to a factor of $\cV$ the metric $\tilde G^{A\bar B}$  is the inverse of $G_{A\bar B}$, 
i.e.~$\tilde G^{A\bar B}=(\cV/2)^{-2} G^{A\bar B}$. 
Having given the metrics in \eqref{decay} 
in terms of a general pre-potential $\cF$, we are free to evaluate them at various points in 
the moduli space parametrized by $t^A$. In the next section we will do that in the 
large volume limit, while section \ref{res_cone} and \ref{multiax} is devoted to a study of the 
axion decay constants on more singular points in the moduli space.

\subsection{Large volume compactifications \label{large_vol}}

Let us now assume that we are in the strict large volume limit in the moduli 
space. This implies that all physical volumes of the two-cycles are larger than 
the square string length $\ell_s^2$. All $\alpha'$ corrections are then suppressed 
and the pre-potential can be approximated by 
\beq \label{F_class}
   \cF^{\rm class} =- \tfrac{1}{3!} \cK_{ABC} t^A t^B t^C\ ,\qquad \quad \cK_{ABC}= \int_Y \omega_A \wedge \omega_B \wedge \omega_C\ ,
\eeq
which is the classical contribution depending on the triple intersections $ \cK_{ABC}$. 
In the large volume limit, $\cV$ given in \eqref{general_K} is simply the geometrical volume of the Calabi-Yau manifold 
and reads \footnote{The volume 
is actually $\cV/8$, but we will keep this normalization for simplicity.}
\beq \label{class_volume}
    \cV = \tfrac{8}{3!} v^A v^B v^C \cK_{ABC}\ .
\eeq
The moduli space metrics in \eqref{C2C4_4d} can be expressed through $\cV$ and 
its derivatives $\cV_A = \partial_{v^A} \cV$ and $\cV_{AB} = \partial_{v^A} \partial_{v^B} \cV$ as
\beq \label{G_AB}
  G_{AB} = - \frac{1}{4} \frac{\cV_{AB}}{\cV} + \frac{1}{4} \frac{\cV_A \cV_B}{\cV^2}  \ ,\qquad \quad 
  \tilde G^{AB} 
  = - 32 \frac{\cV^{AB}}{\cV} + 8 \frac{v^A v^B}{\cV^2}\ ,
\eeq
where $\cV^{AB}$ is the inverse of the matrix $\cV_{AB}$. 
Using this explicit results for the large volume limit,
we will be able to argue that the axion decay constants are sensitive to the
number of axions. This will lead us to conclude that natural scenarios 
contain manifolds with small string-scale cycles. 

Let us now argue that a scenario in the strict $\cN=2$ large volume limit
excludes the possibility of slow roll. As introduced in \eqref{J_exp} the $v^A$ parametrize the string-scale 
size of the two-cycles in the Calabi-Yau space $Y$. They have to take values inside the K\"ahler cone 
in order to ensure that the physical volumes of two- and four-cycles are positive \cite{Wilson}.\footnote{Formally, these conditions translate 
to  $\int_Y J \wedge J \wedge J >0$, $\int_D J \wedge J>0$  and $\int_C J >0$ for all non-trivial divisors $D$ and 
holomorphic curves $C$ in $Y$.} To make this more transparent, we note that, at 
least for toric-projective Calabi-Yau manifolds, we can choose a basis $\omega_A$ such that 
\beq \label{kaehler_cone}
   0 < v^A < \infty\ ,
\eeq 
parametrizes the K\"ahler cone \cite{Wilson, vline}.
In this basis, the large volume limit is obtained when all $v^A > 1$ ensuring that the 
radii $R^A$ in \eqref{J_exp} satisfy $R^A > \ell_s$.
Since the world-sheet instantons contribute
with exponential suppression $e^{- v^A}$
 these $\alpha'$ corrections can be neglected in the above regime. A more detailed 
discussion of the correct choice of basis can be found in, for example, in refs.~\cite{Hosono:1993qy}.

We are now in the position to evaluate the decay constants.
For simplicity, we will assume that all cycles in this limit are 
approximately of the same size $v^A \approx  (R/\ell_s)^2$.
The axion decay constants \eqref{decay} then take the form \cite{SW}
\beq \label{approx_fBCC}
   B_2: \quad \frac{f^2_{AB}}{M_P^2} \cong \left(\frac{\ell_s}{R}\right)^{4} \frac{x_{AB}}{2\pi}\ ,\qquad \qquad C_2\ ,\ C_4\ :\quad  \frac{f^2_{AB}}{M_P^2} \cong g_s^2\, \left(\frac{\ell_s}{R}\right)^{2p} \frac{x_{AB}}{2\pi}\ ,
\eeq
where $p=2$ for the axions of $C_2$ and $p=4$ for the axions of $C_4$. The constants $x_{AB}$ arise from the sum over 
intersection numbers $\cK_{ABC}$ in the metrics $G_{AB}$ and $\tilde G^{AB}$ 
contained in \eqref{decay}. In order to 
get a rough estimate of $x_{AB}$ we note that in the parametrization \eqref{kaehler_cone} of the K\"ahler cone
the intersections are necessarily positive, $\cK_{ABC} \ge 0$ ensuring positivity of the total volume in the 
whole K\"ahler cone.
Inspecting \eqref{G_AB}
 we find that the axion decay constants generically depend on $N$, since we have 
as many axions $(c^A,b^A,\rho_A)$ as volumes $v^A$. For axions of $B_2,C_2$ a rough estimate yields 
$x_{AB} \propto 1/N$ while for axions of $C_4$ we have $x_{AB} \propto 1/N^2$.\footnote{To illustrate 
this result let us concentrate on the first term in $G_{AB}$ in \eqref{G_AB}. For this term 
we evaluate the quotient of  $\cV_{AB}=R^2 \sum_{C} \cK_{ABC}$ and $\cV =\frac{4}{3}R^6 \sum_{D,E,F} \cK_{DEF}$.
We conclude that for each $A,B$ the sums in the numerator runs over significantly more values than
the sum in the denominator.}
 Even though this analysis 
is strictly only valid for Calabi-Yau manifolds with K\"ahler cones \eqref{kaehler_cone} one expects that more exotic examples will 
not significantly alter the conclusion that $x_{AB}$ and hence the axion decay constants $f_{AB}$
scale with $N$. This conclusion rules out large volume scenarios of axion N-flation since the $N$-dependence of $f_{AB}$
will compensate the assistance effect which was of essential importance in achieving slow roll inflation in 
section \ref{rev_N}.

In summary, we note that large volume scenarios make it hard, if not impossible, to implement 
slow roll inflation driven by a large number of axions. This is in accord with the findings
of ref.~\cite{KSS}.  Clearly, a possible conclusion is to relax the 
requirement that all physical cycles are larger than string scale. As we will discuss in the next section,
axions arising on vanishing cycles are natural candidates to realize $N$-flation. Such small cycles 
are not untypical in string theory. However, it is clear that in such scenarios additional stringy corrections 
of the axion decay constants will be of importance and have to be included.

\subsection{Axions from vanishing cycles - The resolved conifold \label{res_cone}}

In this section we discuss axions arising from vanishing cycles of a compact manifold $Y$. 
Since in type IIB string theory axions can arise from the two- and four-forms $B_2,\ C_2$
and $C_4$ a general discussion would include the physics of vanishing two- and 
four-cycles. However, in the following we will exclusively discuss vanishing two-cycles.
The main reason for this restriction is the fact that corrections are significantly better understood 
for lower-dimensional cycles. For two-cycles stringy corrections arise 
from wrapping world-sheets and D1 branes and have been studied intensively 
in the literature \cite{Hori:2003ic,Hosono:1993qy,Ooguri:1996me,RoblesLlana:2006is}.

The standard example for a vanishing two-cycle is the resolved conifold. 
The conifold is a cone over $T^{1,1}$, which is topologically $S^3\times S^2$.
Its metric and harmonic two-form $\omega_2$ are know explicitly 
and the axion decay constants for the R-R axion $c$ of $C_2=\frac{1}{2\pi} c\, \omega_2$
are computed directly \cite{SW}
\beq \label{f_cone}
   \frac{f^2_{\rm cone}}{M_P^2} \cong g_s^2\, \frac{\ell_s^4 \, \Lambda^2}{R^6} \frac{x}{\pi}\ ,
\eeq
where $\Lambda$ is the cutoff regularizing the infinite cone and $x$ is a dimensionless 
number of order $1$. In case we take $\Lambda \approx R$ the result coincides 
with \eqref{approx_fBCC}. Note that this analysis is purely classical and follows 
directly from the geometry of the conifold. In the following we 
will discuss the case where the conifold singularity is resolved by an $S^2\cong \bbP^1$.
We will see that corrections to $f_{\rm cone}$ arise from string world-sheets and D1 instantons wrapped 
around the $S^2$. 

In order to discuss the resolved conifold we recall that in string theory the 
size of the resolution is parametrized by the integral over the K\"ahler form $J$
and the NS-NS two-form $B_2$. 
Explicitly, introducing the complexified K\"ahler structure deformation $t$ as
\beq \label{def-complex_t}
    t = -b + i v= -\tfrac{1}{2\pi} \int_{S^2} B_2 - i J\ ,
\eeq
the size of the $S^2$ is given by $|t|\ell_s^2$. The geometry becomes
singular in the limit $|t|\rightarrow 0$, but remains resolved as 
long as the geometrical volume $v$ or the NS-NS B-field $b$ are 
non-zero. In the following we will study the effects of a small resolution 
by $|t|$ in more detail.
 
 Turning to the corrections to \eqref{f_cone} we recall that the string 
 world-sheet couples to $t$ such that the non-perturbative $\alpha'$ corrections 
 arise through the exponential $e^{i t}$. As discussed in  
 section \ref{four-dim_Lagr} all $\alpha'$ corrections are conveniently encoded by the
 holomorphic pre-potential $\cF(t)$. This pre-potential can be split 
 as 
 \beq \label{pre_split}
     \cF = \cF^{\rm class} +\cF^{\rm pert}+ \cF^{\rm sing} \ ,
\eeq 
where $\cF^{\rm class}$ is 
a cubic classical term given in \eqref{F_class}, and $\cF^{\rm sing}$ contains
the non-perturbative $\alpha'$ corrections from string world-sheets wrapping the vanishing cycle.
The second term $\cF^{\rm pert}$ contains a perturbative  $\alpha'$ correction as 
well as linear and quadratic terms in the K\"ahler moduli. For the discussion of the size
of the axion decay constants we will ignore these contributions, even though the known $\cN=2$
corrections can be straightforwardly included.  
In the case of a conifold singularity the non-perturbative $\alpha'$ corrections 
can be summed up to give a closed expression for $\cF^{\rm sing}(t)$.
The leading pre-potential is thus given by \cite{CdOGP,Hosono:1993qy,GV} 
\bea \label{res_F}
    \cF^{\rm class} &=&\tfrac{i}{3!} (R/\ell_s)^6+\tfrac{i}{2!}(R/\ell_s)^2\, t^2 + \cO (t^3)\ , \nn \\
    \cF^{\rm sing} &=& \tfrac{1}{2i} t^2 \log t +\cO(t^2)\ ,
\eea
for $|t|<1$. Due to the fact that $|t|$ is small we have omitted further terms regular in $t$ in  $ \cF^{\rm sing}$.
Using the quantum corrected pre-potential including the conifold corrections 
we compute the quantum volume \eqref{general_K} as
\beq \label{corr_vol_onecone}
   \cV = \tfrac{4}{3} (R/\ell_s)^6 - 4 (R/\ell_s)^2 v^2 + 2\big( |t|^2 \log |t| + v^2\big) + \cO(t^0) \ ,
\eeq
where $v=(t-\bar t)/2i$ is the volume of the vanishing $S^2$.
It is important to note that for a small resolution $|t|<1$
the logarithmic term in \eqref{corr_vol_onecone} becomes negative. This implies 
that both the second and third term in \eqref{corr_vol_onecone} 
lower the size of $\cV$.

The general form of the leading axion decay constant for the R-R axion $c$ was given in  \eqref{decay}.
It contains the second derivative of the K\"ahler potential \eqref{general_K} which has to be evaluated 
for the resolved conifold pre-potential \eqref{res_F}. 
One derives
\beq\label{f_cone_res}
   \frac{f^2_{\rm cone}}{M_P^2} \cong \frac{g_s^2}{\pi} \frac{ 2 (R/\ell_s)^2- 2 \log |t| -1}{\cV}+\ldots \ ,
\eeq
where $\cV$ is given in \eqref{corr_vol_onecone}. As expected this expression 
correctly reproduces the classical piece \eqref{f_cone}. It also contains the 
leading singular term for $|t|< 1$. We note that, at least to leading singular order,
 the quantum volume in the denominator of \eqref{f_cone_res}
is reduced due to the $\alpha'$ corrections on the conifold singularity, while the expression 
in the numerator increases by a positive term proportional to $-\log |t|$. Hence, the 
size of axion decay constant $f_{\rm cone}$ does not  necessarily decreases if the axion arises from a
small cycle.
This is a desired behavior. It implies that even in the presence of a large number of 
vanishing cycles the axion decay constants are not necessarily small. 
Therefore, we can circumvent a scaling of the decay constants with the 
number of axions as encountered in compactifications with 
all cycles of size $R>\ell_s$ as discussed in section \ref{large_vol}.  
It remains 
to discuss other corrections due to D1 instantons and to estimate the typical 
size of the axion decay constants $f_{\rm cone}$ for the vanishing cycle.

Inspecting \eqref{f_cone_res} one would naively conclude that $f_{\rm cone}$ can be 
made very large by taking the limit $|t|\rightarrow 0$, i.e.~by shrinking the geometrical size 
and also switching off the NS-NS B-field.
However, this will drive us into a regime where the effects of D1 instantons become 
important and correct the axion decay constants as well as the potential significantly. 
This would then imply that the effective action \eqref{simp_act} with potential \eqref{sum-pot}
 for the axion fields will 
not be trustable. To make the coupling to D1 instantons more explicit we introduce the 
fields \footnote{We slightly abuse 
the notation of \eqref{C-expansion} in defining $c$  to also contain the lower R-R form
scalars.}
\beq \label{def-singG}
   G = c - i \frac{|t|}{g_s} = \int_{S^2} (C_2- C_0 B_2) - i  \int_{S^2} e^{-\phi} |B_2-iJ|\ ,
\eeq
where $C_2$ and $B_2$ are the 
R-R and NS-NS two-forms as above. The first integral in \eqref{def-singG}
corresponds to the Chern-Simons action on a D1 instanton, while the second
term is the minimal coupling in the Born-Infeld action.\footnote{In $\cN=1$ compactifications
either the real or imaginary part of $t$ will control the size $|t|$ in the definition of $G$. In section \ref{N=1_ori} we 
show that for O3/O7 orientifolds the B-field $b$ survives in $G$, while in O5 orientifolds
the volume $v$ appears in $G$ \cite{GL1}.}
D1 instantons couple via the exponential 
\beq \label{exp_G}
  Z_{D1}=\exp \big(-i G \big) = \exp \Big(-\frac{|t|}{g_s}   - i c \Big)\ .
\eeq 
In order to 
make sure that there are no significant corrections to the axion decay constants as well 
as no higher harmonics proportional to $Z_{D1}$ 
we have to guarantee that $|Z_{D1}|<1$. This provides a lower bound on the size of $|t|/g_s$ and
hence prevents us from taking $|t|\rightarrow 0$. It will be sufficient to take $|t|$ slightly larger than
$g_s$. In order to get an estimate for the axion decay constants \eqref{f_cone_res} we
also take the typical radius $R$ to be slightly larger than $\ell_s$. 
We thus find 
\beq \label{f<MP}
   f_{\rm cone} \lesssim M_P\ .
\eeq
We  thus conclude that the axion decay constants can be maximally of order Planck scale.
This is not a new result, but rather supports the analysis of ref.~\cite{BDFG} carried 
out in various different string scenarios. 
Equation \eqref{f<MP} implies that we have to make 
use of the assistance effect of many axions as discussed in section \ref{rev_N}. 

Before turning to the multi-axion case let us summarize our strategy. 
As discussed in this section, we will allow for \textit{two} different 
scales in the problem. On the one hand, the radius $R$ sets the size of the 
large cycles, while on the other hand a small parameter $|t|$ sets the size of the 
vanishing cycles. In the following we will 
always take
\beq \label{approx}
   g_s \lesssim 1\ , \qquad \qquad 1\ \lesssim\ |t|/g_s\ ,\qquad \qquad   1 \lesssim\ (R /\ell_s)^2\ .
\eeq 
This choice of hierarchy implies that we can neglect 
world-sheet and D1 instanton corrections on the large 
cycles of radius $R$. On the small cycles of size $|t|$ 
the D1 instantons do not significantly correct the 
metric and hence the axion decay constants. However, 
world-sheet instantons are relevant on these cycles and 
have to be taken into account. Fortunately, in $\cN=2$
set-ups such as the ones discussed in sections \ref{res_cone} and \ref{multiax}, 
these are encoded by the pre-potential and are thus, at least 
in principle, computable by standard techniques such as mirror symmetry.

\subsection{Axions from vanishing cycles - The multi-axion scenario \label{multiax}}

Having discussed the qualitative feature of a single axion from a vanishing 
cycle, we will now turn to the multi-axion case. Recall that due to \eqref{f<MP} 
we always obtain sub-planckian axion decay constants, such that in order 
to implement slow roll inflation as in section \ref{rev_N} a scenario with 
a number of $N$ axions seem unavoidable.
  
In order to introduce set-ups with many vanishing cycles, we divide the 
two-cycles of the compact space $Y$ into two sets.
Let us denote by $\cC_{i} \in H_2(Y)$, $i=1\ldots N$ a set of two-cycles, 
which can be made small without forcing the overall volume of the Calabi-Yau 
manifold to vanish. The remaining cycles in $H_2(Y)$ will be denoted 
by $\cC_{I}$. The two-forms dual to $\cC_{A}=(\cC_i,\cC_I)$
are denoted by 
\beq \label{omega_split}
   \omega_A=(\omega_i,\ \omega_I)\ ,  \qquad \quad i = 1\ldots N\ ,\qquad I = 1\ldots h^{(1,1)}-N\ ,
\eeq
where $h^{(1,1)}$ is the dimension of the second cohomology group $H^{(1,1)}$ of the Calabi-Yau manifold.
The triple intersections $\cK_{ABC}$ defined in \eqref{F_class} are 
evaluated in this basis. As in section \ref{res_cone} we will introduce two scales and 
assume that all large cycles are of
approximate radius $R\gtrsim \ell_s$ and all small cycles are of 
size $|t|$, i.e.~we demand
\beq \label{valpha_same}
   v^I \approx (R/\ell_s)^2 \gtrsim 1\ ,\qquad \qquad |t^i| \approx |t| <1 \ .
\eeq
As an estimate of complicated final expressions we will express them 
as functions of $R$ and $|t|$ as well 
as the number $N$ of axions only.

The leading classical axion decay constants for the 
R-R axions $c^i$ have been derived in section \ref{large_vol}. They are 
obtained by taking the derivatives of the classical volume 
\eqref{class_volume}.
The resulting classical metric $G_{ij}$ given in \eqref{G_AB} 
is inserted into the general expression \eqref{decay} for the decay constants.
As in the example of the resolved conifold discussed in section \ref{res_cone}
it will be crucial to evaluate the quantum corrections to this classical result.
To make this explicit, this would force us to specify the type of singularities
 we are considering and thus requires further information about the geometry of the set-up.
In this paper we will only discuss a simple toy model of $N$ conifolds and 
it would be desirable to extend this analysis in future work. However, the general 
way to proceed should be as follows. In the split \eqref{pre_split} of the pre-potential $\cF$
the contribution $\cF^{\rm sing}$ will only depend on the small moduli $t^i$ and 
should be derived for a given geometry. This singular geometry can be 
considered to be local or non-compact and later embedded into a compact 
space with cut-off given by the radius $R$ of the large cycles.
Given a pre-potential \eqref{pre_split} the leading axion decay constants are 
straightforwardly computed by evaluating
\beq \label{general_f_cone}
     \frac{f^2_{ij}}{M_P^2} =\frac{g_s^2}{ \pi} G_{i \bj}\ ,  
\eeq
where $G_{i \bj}=\partial_{t^i}\partial_{\bar t^j} K$ is the K\"ahler metric to
the K\"ahler potential \eqref{general_K}.

Let us now exemplify a multi-axion scenario with a 
toy model admitting $N$ resolved conifold singularities. 
To derive the axion decay constants we investigate the pre-potential 
\beq \label{toy_F_N}
   \cF = \tfrac{i}{3!} (R/\ell_s)^6 +  \tfrac{i}{2} \sum_{i=1}^N (t^i)^2 \big[ (R/\ell_s)^2   -  \log t^i \big] + \ldots\ ,
\eeq
which is the straightforward generalization of the single conifold pre-potential \eqref{res_F}.
It can be viewed as describing a toy model of $N$ conifold singularities 
in a compact Calabi-Yau manifold with simple intersection numbers. 
We have assumed that the non-trivial intersection 
numbers of the large cycles of radius $R$ with the vanishing cycles of size $|t^i|$ are given by
$\cK_{I ij} \approx - \delta_{ij}$, while all other intersections of the large cycles 
are of order one $\cK_{IJK}\approx \cO(1)$ or vanish. 
The quantum corrected volume \eqref{general_K} is computed 
to be 
\bea \label{toy_V_N}
  \cV &=& \tfrac{4}{3}  (R/\ell_s)^6  -4 (R/\ell_s)^2  \sum_i (v^i)^2 + 2  \sum_i \big( |t^i|^2 \log |t^i| +(v^i)^2 \big)+ \ldots \nn \\
     &\approx&\tfrac{4}{3}  (R/\ell_s)^6  - 4 N (R/\ell_s)^2 v^2 + 2 N \big(|t|^2 \log |t|+v^2 \big) + \ldots\ .
\eea
where in the second line we have used  \eqref{valpha_same}.
This expression should be compared to the case of a single conifold \eqref{corr_vol_onecone}.
The axion decay constants are computed using the general expression \eqref{general_f_cone}. Inserting 
the pre-potential \eqref{toy_F_N} we find that the leading contributions to $f_{ij}^2 = \delta_{ij} f^2_i$ are diagonal with
\beq \label{sing_axion_decay_ij} 
  \frac{f^2_{i}}{M_P^2} =\frac{g_s^2}{\pi} 
  \frac{2(R/\ell_s)^2- 2 \log |t| -1}{ \cV} + \ldots\ .
\eeq
We can now draw the conclusion already indicated in the previous section. Even though
we are dealing with a set-up with a possibly large number $N$ of small cycles, the axion 
decay constants can be close to the Planck scale for appropriate values of $R$, $t$ and $g_s$.

To illustrate \eqref{sing_axion_decay_ij} let us consider a numerical 
example. Let us assume that we have $100$ large cycles of volume 
$(R/\ell_s)^2 \approx 1.44$ and to each there are coupling $8$ different 
small cycles of volume $|t| \approx 0.31$
with coupling $(R/\ell_s)^2 (t^i)^2$ in \eqref{toy_F_N}.
This gives $N=800$ axions.
At string coupling $g_s=0.25$ one evaluates 
\beq \label{fi01}
    f_{i} \approx 0.1\, M_P\ ,
\eeq
 at a string-frame volume $\cV\approx 18.6$ and Einstein-frame volume $V_E = g_s^{-3/2} \cV \approx 148$ in units of $\ell_s$.
 We also evaluate 
 \beq
    \frac{|t|}{g_s} \approx 1.25\ ,\qquad \frac{(R/\ell_s)^2}{g_s} \approx  5.7\ . 
 \eeq

To end this section let us also comment on the corrections due to D-brane instantons
to the axion decay constants $f_{ij}$. 
In particular, D1 instantons will 
introduce a dependence of $f_{ij}$ on the axion fields $c^i$ themselves. This 
is clear from the fact that D1 instantons on a cycle $\cC_i$
couple to the complex coordinates $G^a$ defined as  \footnote{We slightly abuse 
the notation of \eqref{C-expansion} in defining $c^i$ to also contain the lower R-R form
scalars.}
\beq \label{def-Ga}
 G^i = c^i - i \frac{|t^i|}{g_s} = \int_{\cC_i} (C_2- C_0 B_2) - i  \int_{\cC_i} e^{-\phi} |B_2-iJ|\  ,
\eeq
which is the obvious generalization of \eqref{def-singG}. The 
corrections are proportional to the exponential of the $G^a$ as
in \eqref{exp_G}.
In complete analogy to the discussion in section \ref{res_cone} 
we want to make sure that these D-brane corrections are parametrically 
small and can be neglected in the analysis of the axion decay constants. 
Therefore, we demand that each $|t^i|/g_s$ is 
larger than unity, such that $|Z_{D1}|\ll 1$. In other words, 
we will demand that each $t^i \approx t$ satisfies the constraint
\eqref{approx}. Again, we have two scales parametrizing our models; 
one parameter $R$ setting the size of the large cycles, and another much 
smaller parameter $|t|$ associated to the stringy volume of the vanishing cycles.

\section{Axion potentials in type IIB  string theory \label{axion_potential}}

In the previous section we studied the kinetic terms of R-R axions for 
type IIB Calabi-Yau compactifications. Most of our analysis 
focussed on axions $c^a$, $a=1\ldots N$ arising from the two-form $C_2$ 
integrated over vanishing cycles in the Calabi-Yau manifold. In this 
section we will discuss the non-perturbative effects which lead 
to the generation of a non-vanishing scalar potential
for the fields $c^a$.  We will identify sources which generate an  
axion potential of the form
\beq \label{ax_recall}
     V_{\rm eff}(\theta^a) = C + \sum_{a=1}^N \Lambda^4_a \big(1- \cos \big[\mu^a\, \theta^a/f_a \big] \big) \ ,
\eeq
which was already given in  \eqref{sum-pot}. In this effective potential,
the $\theta^a$ are the canonically normalized axions \eqref{def-theta} and $f_a$ are the diagonalized axion 
decay constants. 

To make the discussion of axion potentials more  
explicit we will further break the supersymmetry of the $\cN=2$
scenarios of section \ref{axion_decay} to $\cN=1$. This can be done by including 
background fluxes, space-time filling D-branes or orientifold planes \cite{review_flux}. If these 
additional sources are appropriately chosen they only preserve half
of the eight supercharges in four dimensions. It is well-known, that 
the effective action of the resulting four-dimensional supergravity
theory can be described in terms of a set of $\cN=1$ characteristic functions.
The kinetic terms 
for a set of chiral multiplets containing the complex scalars $M^n$ 
are encoded by the metric $K_{n \bar m} =\partial_{M^n}\partial_{\bar M^m} K$,
which is the second derivative of a K\"ahler potential $K$.
Supersymmetry also implies that the effective 
scalar potential can always be written as 
\beq \label{scalarpot}
  V = e^{K/M_P^2}(K^{n\bar m} D_{n} W \overline{D_{m} W} - 3|W|^2/M^2_P) + (\R f)^{kl} D_{k} D_l\ .
\eeq
Here $W$ is the superpotential holomorphic in $M^I$ and 
$D_n W = \partial_{M^n} W +(\partial_{M^n} K) W$ denotes its K\"ahler covariant 
derivative. The second term in \eqref{scalarpot} arises if some of the 
scalars are gauged. It depends on  the non-trivial D-terms $D_k$
as well as the real part of the gauge-kinetic coupling function $f_{kl}$.
In the following we will discuss sources for a superpotential $W$ which 
inserted into the potential \eqref{scalarpot} induces an axion potential 
of the form \eqref{ax_recall}.

\subsection{Corrections due to D1 instantons \label{D1_corrections}}

In this section we will discuss a first set of corrections 
which can induce a non-trivial axion potential of the form 
\eqref{ax_recall}. More precisely, we will study an effective 
potential arising from D1 instantons 
wrapped around small cycles in the Calabi-Yau manifold.

In order to study the effects of D-instantons we will focus entirely on the 
superpotential $W$. Recently, much effort has focussed on the study of 
D3 instantons wrapped on four-cycles in O3/O7 orientifold and F-theory compactifications 
as reviewed, for example, in refs.~\cite{review_flux}.
These instantons can, under certain well-known conditions \cite{Witten1}, induce  
a superpotential of the form
\beq \label{D3-corrections}
  W_{\rm D3} = \sum_{\alpha}  A_\alpha \ e^{i n_\alpha^{\, \beta}\, T_\beta}\ .
\eeq
In this expression $T_\beta$ are the moduli containing 
the K\"ahler structure deformations $v^\beta$ defined in \eqref{J_exp}. 
In \eqref{D3-corrections} the functions $A_\alpha$ generically
depend on other fields in the spectrum, while 
$n_\alpha^{\ \beta}$ is a constant matrix parametrizing the wrapping numbers of 
the D3 instantons.
In the work of KKLT \cite{KKLT} the corrections \eqref{D3-corrections} have been 
used to stabilize the moduli $T_\alpha$ at large volume $v^\alpha>1$. Moreover,
in specific set-ups the corrections \eqref{D3-corrections}, together with a 
superpotential due to background fluxes, are shown to be sufficient 
to stabilize all moduli in the model \cite{Denef:2005mm}. Typically,
such examples have no R-R two-form scalars in the spectrum, since these 
have been projected out in the $\cN=2$ to $\cN=1$ reduction.
In order to realize the scenarios of section \ref{decay} we 
thus have to focus on a more general set of examples
which contain R-R two-forms in the spectrum.

Examples admitting R-R two-form axions have been studied, for example, in 
refs.~\cite{Lust:2006zh,TG}. In such O3/O7 orientifolds the fields $G^a$ defined in \eqref{def-Ga}
arise as chiral multiplets in the spectrum. 
Based on the earlier works \cite{Witten3,Ganor}, it 
has been argued in \cite{TG} that a dependence on $G^a$ can appear through the 
pre-factors $A_\alpha$ in \eqref{D3-corrections}. 
More precisely, this dependence arises through generalized theta-functions
$ \Theta_\alpha(\tau,G)$ each being a power series in the exponentials $e^{i\tau}$ and $e^{-iG^a}$. 
The superpotential \eqref{D3-corrections} can be written as 
\beq \label{D3_superpotential}
   W_{\rm D3} = \sum_{\alpha}\hat A_{\alpha} \ \Theta_\alpha(\tau,G) \ e^{i n_\alpha^{\, \beta}\, T_\beta}\ ,
\eeq
with new coefficient functions $\hat A_{\alpha}$ independent of $\tau,G^a$.\footnote{An explicit compactification, 
where such a superpotential is induced, can be found in ref.~\cite{Florea:2006si}.}
For our purposes it will be sufficient to consider the first few terms in $\Theta(\tau,G)$. 
For example, consider a set-up with one field $T$ coupling to a set of moduli $G^1,\ G^2,\ $etc.
Since $\Theta(\tau,G)$ generically starts with a constant, a candidate superpotential is of the 
form
\beq \label{D3_superpotential_expand}
   W_{\rm D3} =  e^{i T} + e^{i\tau} \big[e^{-iG^1} + e^{-iG^2}+\ldots  \big] \ e^{i T}\ ,
\eeq
where the pre-factors were taken to be of order one.
Note that 
due to the presence of the exponential factor involving $T_\beta$ the $G^a$ dependence 
through $\Theta_\alpha(\tau,G)$ would be sub-leading if contributions entirely due to D1 instantons are 
present. This is the case since $\I T_\alpha>1$ parametrizes large volumes 
of four-cycles in a self-consistent analysis. However, as can be seen by analyzing the
F-theory underlying the $O3/O7$ orientifold, there is no calibration which can support 
D1 instantons alone \cite{Witten3,Ganor,TG}.

Let us now turn to potentials for the 
R-R two-forms scalars entirely generated by effects of D1 instantons.
These couple to the complex coordinates $G^a$ defined in \eqref{def-Ga}.
In analogy to the discussion of D3 instantons one expects these 
to generate a superpotential of the form 
\beq \label{D1-corrections}
  W_{\rm D1} = \sum_a B_a \ e^{-i m_a^{\, b}\, G^a}\ .
\eeq 
This form is not surprising since non-perturbative corrections are 
necessarily weighted by the exponential of the action of the corresponding 
instanton. Hence, as in \eqref{exp_G} D1-instantons contributions on vanishing 
cycles contain an exponential factor of $G^a$. A study of explicit examples will 
reveal if the coefficient functions $B_a$ are indeed non-vanishing for the 
D-brane configurations under consideration. 

A detailed investigation of the superpotential \eqref{D1-corrections} has 
been carried out for type I string theory on a Calabi-Yau manifold in \cite{Witten2}.
Type I strings can be viewed as an orientifold set-up with O9 planes
and D9 branes. In this case the conditions for a non-vanishing 
superpotential \eqref{D1-corrections} are known explicitly. 
One expects that the analysis of \cite{Witten2} can be extended to 
the other orientifolds such that a superpotential 
\eqref{D1-corrections} will be generated for part or all of the moduli $G^a$.
For example, explicit computations of such a D1 superpotentials including multi-instanton 
contributions has been recently carried out in \cite{ABK}.

\subsection{Gaugino condensates on D5 branes \label{D5_gaugino}}
 
In the following we comment on an interesting further 
source, which induces the desired axion potentials. More precisely, 
let us consider  space-time filling D5 branes which wrap the small 
cycles in the compact Calabi-Yau manifold. Such branes can be 
supersymmetric and consistently included in orientifold set-ups with O5 planes.
In the presence of appropriate orientifold planes the 
tadpoles can be cancelled yielding a stable configuration 
\cite{review_flux}. Alternatively, one might try to include pairs of D5 and anti-D5 
branes wrapping two-cycles in the same homology class \cite{Aganagic:2006ex}.
Even though such configurations are only meta-stable and will break supersymmetry, they can be 
sufficiently long lived to accommodate our universe. 

For concreteness, let us first focus on a stack of $M$ D5 branes wrapping a small 
$S^2$ resolving the conifold singularity as in section \ref{res_cone}. 
The gauge theory on the space-time filling D5 branes is a pure 
$U(M)$ Yang-Mills theory with $\cN=1$ supersymmetry. 
At low energies this gauge theory will be strongly coupled 
leading to gaugino condensation.
The gauge coupling is given by the complex field  
$G = c - |t|/g_s$ already defined in \eqref{def-singG}. 
The effective theory 
for the gaugino condensate $S$ admits a superpotential of 
Veneziano-Yankielowicz form
\beq \label{VY_potential}
   W_{\rm VY} = -i G\, S + \tfrac{1}{2\pi i} M\, S\big(\log(S/\Lambda^3_0)-1 \big)\ ,
\eeq
where $\Lambda_0$ is the cutoff scale. The extrema of $W_{\rm VY}$
correspond to the vacua of the gauge theory. Eliminating $S$ the 
effective superpotential is of the form 
\beq   \label{D5_superpotential}
   W_{\rm D5} = \Lambda_0^3  e^{-i \frac{G}{M}}\ .
\eeq
The superpotential has a dependence on the field $G$ similar to the one 
arising from D1 instantons \eqref{D1-corrections}. However, there is 
a crucial factor of $1/M$ appearing in the exponential. In the axion potential \eqref{ax_recall}
we thus identify $\mu = 1/M$ where $\mu$ is appearing in the cosine multiplying the 
canonically normalized axion. 

\begin{figure}[!ht]
\leavevmode
\begin{center}
\includegraphics[height=4.1cm]{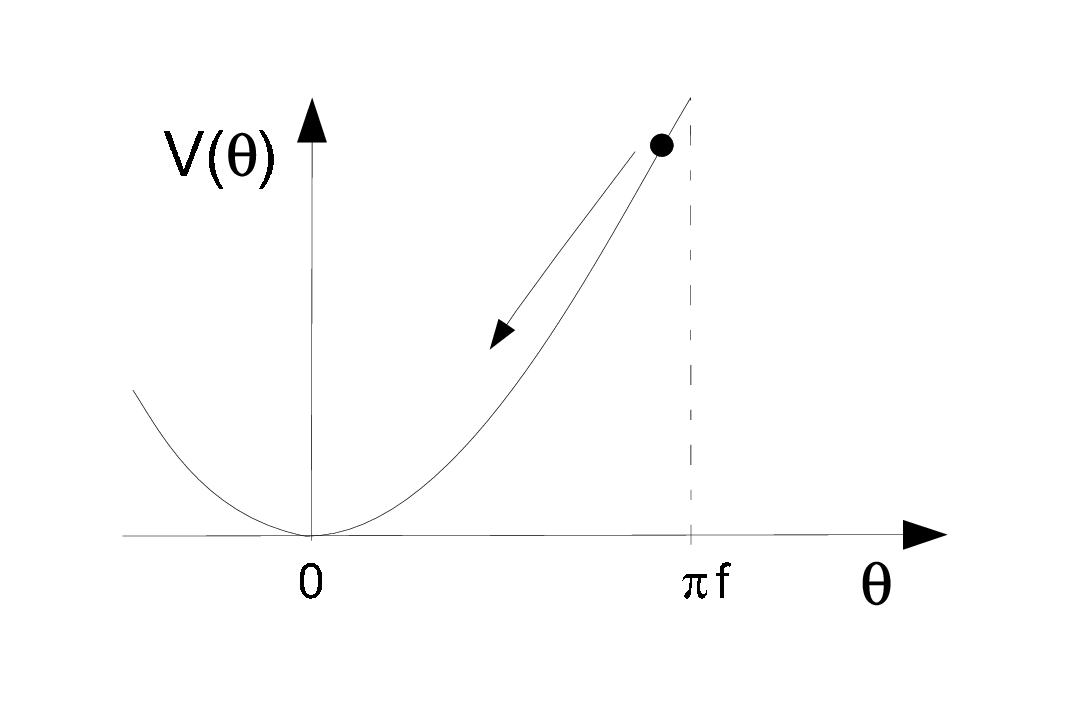} 
\end{center}
\vspace*{-.5cm}
\caption{\small Potential for one axion field $\theta$ with $M>3$. }
\label{Potential2}
\end{figure}

It is important to note that such a result has 
to be treated with care. The factor $1/M$ should not spoil the $2\pi$ periodicity 
of the axions. This can be achieved by replacing the potential \eqref{ax_recall} with
\beq \label{2pi_period}
   V_{\rm eff} =C+ \sum_a \Lambda_a^4 \mathop{\text{min}}_{n^a \in \bbZ}\Big(1-\cos\Big[ \frac{(\theta^a/f_a) + 2\pi n^a}{M^a}\Big] \Big)\ ,
\eeq
for gaugino condensates of  an $U(M^a)$ gauge theory on the $a$th stack of D5 branes. 
That such a form is indeed obtained by analyzing the confining gauge theories using string duality
was explicitly shown in refs.~\cite{Heckman:2007ub}. The form \eqref{2pi_period} implies that
the field-range of the canonically normalized axions $\theta^a$ will remain limited 
by the value of the axion decay constants alone. Considering $c^a$ and $\theta^a$
in the intervals \eqref{intervals}
the potential \eqref{2pi_period} is again identical 
to \eqref{ax_recall} with $\mu^a = 1/M^a$.
Remarkably, independent of the value of the 
axion decay constants the factors $1/M^a$
can push the smooth maximum of the cosine out of the accessible field range of 
$\theta^i$ (compare figure \ref{Potential} with figure \ref{Potential2}).  For $M^a>3$ the potential 
becomes approximately quadratic in the complete field range  \eqref{intervals} of
$\theta^i$ and we can apply 
equations \eqref{eps_eta_quad} and \eqref{Nrhor_quad} for all values of $\theta^i$.

Since this seems an interesting opportunity to built explicit models for chaotic inflation, 
let us briefly comment on the computation of \eqref{2pi_period} in 
refs.~\cite{Aganagic:2006ex,Heckman:2007ub}. The authors consider 
stacks of several D5 (and anti-D5 branes) in the same homology class on 
a small two-cycle. 
This configuration can be pushed through a conifold 
transition. In this process the branes get replaced by three-form 
fluxes on small three-spheres $S^3$ with complex structure modulus 
$S$ replacing the gaugino bilinear \cite{Vafa:2000wi}. In the dual picture 
the superpotential and scalar potential are entirely determined by the local geometry. 
It can be given in terms of the dual pre-potential $\hat \cF(S)$ as
\beq \label{flux_pot}
   W = \int G_3 \wedge \Omega(S) = - i G\, S + M\, \partial_S \hat \cF(S)\ ,
\eeq
where $G_3$ is the complex NS-NS and R-R three-form flux. 
In \eqref{flux_pot} the pre-potential takes the form
$\hat \cF = - \frac{1}{2\pi i} S^2 \log( S/\Lambda_0^3) + f_{\rm geom}(S)$, with $f_{\rm geom}(S)$ 
encoding further corrections to \eqref{VY_potential} given by the local geometry.
Eliminating $S$ in the local vacuum allows 
to explicitly compute \eqref{2pi_period} and determine $\Lambda_a$ \cite{Heckman:2007ub}.
Note that in order to fully embed these local results into a global scenario
further work remains to be done. However, combining available techniques 
for geometric transitions with the analysis of the axion decay constants 
and scales of section~\ref{axion_decay} provides interesting possibilities 
to establish explicit scenarios \cite{inprogress}.

Before moving on to the $\cN=1$ orientifold scenarios, 
let us stress the points we still need to address in 
order to make our scenarios consistent.
Having analyzed the axion 
decay constants encoding the kinetic terms and 
the scalar potential due to non-perturbative D-brane effects,
it remains to discuss  two important issues:\\ 
(a) It has to be ensured that
all non-axionic moduli are stabilized at the desired scales.\\
(b) It has to be true that the axions are the only fields which are relevant during 
inflation. In particular, the non-axionic scalar 
partners in the supermultiplets of the axions should not interfere with 
their dynamics during inflation.\\
Both of these requirements are hard to address in general and are not
necessarily satisfied in many string compactifications \cite{KSS}.
First steps in establishing a consistent scenario for a class of $\cN=1$ 
orientifold compactifications will be made in the next section.

\section{Axion Inflation in $\cN=1$ orientifolds \label{N=1_ori}}   

In the following we embed the scenario outlined in previous sections
into an $\cN=1$ orientifold compactification. 
We begin with the study of the four-dimensional 
effective $\cN=1$ theory obtained by compactifying 
Type IIB string theory on a Calabi-Yau orientifold with O3/O7 planes in section \ref{four-dim_N=1}. 
The $\cN=1$ data are expressed as a function of the underlying $\cN=2$ pre-potential and 
hence inherit the $\alpha'$ corrections discussed in section \ref{axion_decay}. 
This will be illustrated for a toy model with $N$ conifolds in section \ref{N_conifolds}.
Combined with the superpotentials of section \ref{axion_potential} 
an approximate evaluation of the potential allows us to comment on axion inflation
in these $\cN=1$ scenarios.

\subsection{The four-dimensional $\cN=1$ effective theory \label{four-dim_N=1}}   

Let us discuss the $\cN=1$ effective theory in more detail. 
As above, the compact Calabi-Yau manifold is denoted by $Y$. 
In order to define the orientifold projection we demand that $Y$ admits 
a holomorphic, isometric involution $\sigma$. The full orientifold 
action is given by 
\beq \label{Oaction}
  \cO = (-)^{F_L} \Omega_p \sigma^*, \qquad \quad \simga^* J =J\ , \qquad \quad \sigma^* \Omega =- \Omega
\eeq
where $\Omega_p$ is the world-sheet parity, $F_L$ is the left moving fermion number
and $\sigma^*$ is the pull-back of $\sigma$ acting on forms. In \eqref{Oaction} we have also displayed 
the action of $\sigma$ on the  K\"ahler form $J$ and holomorphic three-form $\Omega$ of  
$Y$. $J$ is invariant due to the fact that $\simga$ is holomorphic and isometric. 
The negative sign for the action on $\Omega$ implies that we are considering 
scenarios with $O3$ and $O7$ planes.

In order to determine the $\cN=1$ spectrum we note that 
$\sigma$ splits the cohomologies into positive and 
negative eigenspaces. 
In particular, the cohomology of two-forms splits as 
\beq \label{H^2split}
   H^{(1,1)} = H^{(1,1)}_+ \oplus H^{(1,1)}_-\ ,
\eeq
with dimensions $h^{(1,1)}_+$ and $h^{(1,1)}_-$. Accordingly, the basis of $H^{(1,1)}$
can be split into $\omega_A = (\omega_\alpha,\omega_a)$ with $\omega_\alpha$
being a basis of the positive eigenspace in \eqref{H^2split} and $\omega_a$ being a 
basis in the negative eigenspace of \eqref{H^2split}. 
Note that the K\"ahler form $J$ lies in the positive eigenspace 
of \eqref{H^2split} as indicated in \eqref{Oaction}. In contrast, the NS-NS B-field 
has to transform as $\simga^* B_2 = - B_2$ in order to remain in the spectrum and thus 
lies in the negative eigenspace of \eqref{H^2split}. Hence, in the basis 
$\omega_A=(\omega_\alpha,\omega_a)$
we have to expand
\beq \label{JBexpand}
   J = v^\alpha \omega_\alpha \ , \qquad \quad B_2 = b^a \omega_a\ , \qquad \quad t^a = -b^a\ ,\qquad t^\alpha = iv^\alpha\ ,
\eeq 
where $t^A=(t^\alpha,t^a)$ are the complexified K\"ahler structure deformations 
introduced in \eqref{def-t}.
We find that the negative two-cycles dual to $\omega_a$ have vanishing geometric 
volume $v^a=0$, but are supported by a non-trivial B-field. 
The positive two-cycles dual to $\omega_\alpha$
have geometric volume $v^\alpha$ and vanishing B-field.

The four-dimensional $\cN=1$ effective theory obtained by compactifying
on the Calabi-Yau orientifold $Y/\sigma$ was derived in \cite{GL1,GL2}. 
It was shown in \cite{GL1} that the scalar modes arising from $B_2$ as
well as $C_2,C_4$ combine together with the K\"ahler structure deformations 
into chiral $\cN=1$ multiplets. The complex scalars in the chiral multiplets 
are the dilaton-axion $ \tau=C_0 + ie^{-\phi}$, 
the purely axionic fields $G^a$ already given in \eqref{def-Ga} and 
the complexified K\"ahler moduli $T_\alpha$. Explicitly, we define
\beq \label{def-T}
    G^a = c^a + ie^{-\phi} \R\, t^a \ ,\qquad \quad 
   T_\alpha =-\rho_\alpha + i e^{-\phi} \R\, \cF_{\alpha} \ ,
\eeq
where $\R\, t^a=-b^a$ and the axions $c^a$ and $\rho_\alpha$ are defined as
\beq \label{axions_eta_rho}
    c^a =\tfrac{1}{2\pi}\int_{\cC_a} C_2 - C_0 B_2\ ,\qquad\qquad  
    \rho_\alpha =\tfrac{1}{2\pi} \int_{\tilde \cC^\alpha} C_4 - B_2 \wedge C_2 + \tfrac12 C_0 B_2 \wedge B_2\ .
\eeq 
In \eqref{axions_eta_rho} the cycles $\cC_a \in H_2^-$ are dual to $\omega_a$ and $\tilde \cC^\alpha \in H_4^+$ are 
four-cycles transforming with a positive sign under $\simga^*$. At the end of section \ref{multiax},  
the fields $G^a$ have been identified as 
the correct couplings to D1 instantons for  two-cycles of vanishing geometrical volume. 
The coordinates $T_\alpha$ provide the correct couplings to D3 instantons 
wrapping four-cycles in $Y$. 

Due to the $\cN=1$ supersymmetry of the four-dimensional effective 
action the metric on the field space spanned by $\tau, G^a$ and $T_\alpha$
is necessarily K\"ahler. This implies that it takes the form $G_{I\bar J} = \partial_{M^I} \partial_{\bar M^J} K_{\rm q}$, where 
$M^I=(\tau,G^a,T_\alpha)$ are the complex coordinates and $K_{\rm q}(M,\bar M)$ is the K\"ahler potential. 
The K\"ahler potential for the complex scalars $M^I$ is
shown to be \cite{GL2,TG}
\bea \label{Kqgeneral}
 K_{\rm q}(\tau,G,T) = - 2 \ln \big[i e^{-2\phi} \big( 2(\cF -\bar \cF)-(\cF_A +\bar \cF_A)(t^A-\bar t^A) \big) \big]\ .
\eea
It is important to note that despite of the obvious similarity to \eqref{general_K} the K\"ahler 
potential $K_{\rm q}$ has a more complicated functional dependence on its complex
coordinates $M^I$. 
For a general pre-potential $\cF$ it is impossible to explicitly write $K_{\rm q}$ as the function of $\tau,G^a,T_\alpha$.
This is due to the fact that one would need to express  $e^{-2\phi} \cV$ as a function of $e^{-\phi} \R\, t^a$ and $e^{-\phi} \R\, \cF_{\alpha}$
 appearing in the $\cN=1$ coordinates \eqref{def-T}. This functional dependence is non-polynomial 
 and can only be determined explicitly in very specific 
 examples.\footnote{This is equivalent to the problem of solving the attractor equations 
 for $\cN=2$ black holes.}
 Nevertheless, one can derive the K\"ahler metric
 by using the underlying $\cN=2$ special geometry \cite{GL2} or the work of Hitchin \cite{Hitchin} as
 done in \cite{BG}. Also the derivatives of the K\"ahler potential \eqref{Kqgeneral} are known as a function of the
pre-potential \cite{GL2}. The K\"ahler metric and its inverse as well 
as the first derivatives of the K\"ahler potential are summarized in appendix~\ref{N=1_recall}.

 Before moving on to the $\cN=1$ axion decay constants let us stress again 
 that $K_{\rm q}$ still contains the $\alpha'$ corrections inherited 
 from the underlying $\cN=2$ theory. It does, however, not  depend on the 
axions $C_0,c^a,\rho_\alpha$. This can be traced back to the fact that no corrections 
due to D-branes are included in the expression \eqref{Kqgeneral} which is in accord with the
discussion in sections \ref{res_cone} and \ref{multiax}. Clearly, in addition 
to the inherited $\cN=2$ corrections one expects further $\cN=1$
corrections to appear. However, it seems unlikely that these 
will cancel the $\cN=2$ effects. We thus have some confidence 
that the appearance of large axion decay constants is not 
restricted to the $\cN=2$ set-ups outlined in section  \ref{axion_decay}.

Let us discuss the axion decay constants 
for the R-R two-form axions $c^a$ in \eqref{def-T} in more detail.
Using $K_{\rm q}$ these are simply given by the K\"ahler metric 
\beq \label{N=1axion_decay}
   \frac{f^2_{ab}}{M_P^2} = 2\, \partial_{G^a} \partial_{\bar G^b} K_{\rm q}\ .
\eeq
As a consistency check we can derive $f_{ab}$ for the classical 
pre-potential \eqref{F_class}. This was done in ref.~\cite{GL1} (appendix C.1) and reproduces
the leading classical result discussed in section \ref{large_vol}.
In the special cases such as when $h^{(1,1)}_+=1$, i.e. in case there is 
only one $T_1 \equiv T$, the computation of the classical $K_{\rm q}$ 
simplifies significantly and \eqref{Kqgeneral} reduces to  
\beq \label{KqoneT}
    K_{\rm q}= -  \ln i(\tau-\bar \tau) - 3 \ln i \Big(T-\bar T-\frac{C_{ab} (G-\bar G)^a (G-\bar G)^b}{2(\tau -\bar \tau)} \Big)\ ,
\eeq
where $C_{ab}$ is a positive definite integral matrix given by the triple 
intersection $C_{ab}=- \int \omega \wedge \omega_a \wedge \omega_b$, with $\omega \in H^{(1,1)}_+$
corresponding to the one modulus $T$. It is now straightforward to compute $f_{ab}$ and 
compare the result with the $\cN=2$ counterpart of section \ref{large_vol}.

Recall that we found in sections \ref{res_cone} and \ref{multiax} that 
in order to obtain large axion decay constants we need 
to move close to singular points in the moduli space. In the vicinity of a singularity
the $\alpha'$ corrections in the K\"ahler potential $K_{\rm q}$ become important
and have to be included. 

\subsection{A simplistic $N$ conifold toy model \label{N_conifolds}}

In this last section we will study an orientifold scenario
of a simplified toy model with $2N$ resolved conifold singularities. 
The aim of this section is to illustrate that the $\cN=1$ effective theory 
can be evaluated explicitly including the simple logarithmic $\cN=2$
corrections near the singularities. This allows to illustrate some of the 
features necessary to model $N$-flation. 
However, the reader should not consider this as an 
explicit construction, but rather as support of our general 
arguments that the outlined scenarios provide a promising
possibility to model axion inflation.

Let us start with a compact geometry with a 
set of $2N$ conifold singularities resolved by two-spheres $S^2_i,S^2_{-i}$, 
where $i=1\ldots N$. As in section \ref{multiax}, 
the blown-up spheres are supported by a geometric volume and NS-NS B-field.
We assume that there is an orientifold projection mapping \footnote{The local geometry 
of this set-up is similar to the T-dual of the O5 orientifold geometries discussed, for 
example, in ref.~\cite{ABK}.}
\beq
   \sigma S^{2}_ i  = S^2_{-i}\ ,\qquad \quad i=1\ldots N\ .
\eeq
The two-forms associated to the two-cycles $(S^{2}_i,S^2_{-i})$ are denoted by
$(\tilde \omega_i,\tilde \omega_{-i})$. To obtain invariant and anti-invariant forms
these are combined as
\beq
   \omega_i^+ = \tilde \omega_i +\tilde \omega_{-i}\ ,\qquad \qquad  \omega_i^-= \tilde \omega_i -\tilde \omega_{-i}\ ,
\eeq
with $\omega_i^\pm$ transforming with a plus or minus sign under the orientifold involution $\sigma^*$.
The corresponding special coordinates are likewise given by
\beq \label{t+-}
   t^i_+= \tfrac{1}{2} (t^i + t^{-i}) = i v^i,\qquad \quad t^i_-=\tfrac{1}{2} (t^i - t^{-i}) = -b^i\ ,\qquad \quad t_{\rm R} = i (R/\ell_s)^2\ ,
\eeq
where $t^{i},t^{-i}$ are the complexified K\"ahler moduli 
corresponding to the resolved conifolds and $t_{\rm R}$ parametrizes 
the larger cycles of radius $R$. 
Note that in \eqref{t+-} we have already applied the orientifold constraint \eqref{JBexpand}
to express $t^i_\pm$ through the B-field and K\"ahler form alone. 
In other words, the orientifold projection enforces that the volumes
of $S^i$ and $S^{-i}$ are identical, $v^i=v^{-i}$, while the B-fields
have opposite sign, $b^i=-b^{-i}$.

In order to derive the metric on the $\cN=1$ field space, we will consider the simple 
pre-potential  \eqref{toy_F_N} written as
\beq
   \cF  =- \tfrac{1}{3!}\, t_{\rm R}^3 + \tfrac{1}{2!}   \sum_{i=-N}^N (t^i)^2 \big[t_{\rm R} -i \log t^i \big]\ ,
\eeq
where the sum does not include the term with $i=0$.
Using the coordinate transformation \eqref{t+-}, $\cF$ can be expressed in terms of $t^i_\pm$, $t_{\rm R}$, 
and hence $v^i, b^i$ and $R$,
as
\bea \label{orientifold_pre}
   \cF & =& -\tfrac{1}{3!} t_{\rm R}^3 + t_{\rm R}  \sum_{i=1}^N \big[(t_+^i)^2 + (t_-^i)^2 \big]   \\
      && - \tfrac{i}{2}\sum_{i=1}^N (t_+^i + t_-^i)^2 \log (t_+^i + t_-^i) - \tfrac{i}{2}\sum_{i=1}^N (t_+^i - t_-^i)^2 \log (t_+^i - t_-^i)\ . \nn
\eea
This explicit ansatz for $\cF$ should be understood as our main simplification 
in the study of these $\cN=1$ orientifold models with resolved conifold singularities. 
In general, there will be further perturbative and non-perturbative corrections 
to the $\cN=1$ K\"ahler potential depending on $b^i, v^i$ and $R$ 
which are only suppressed in certain regimes of the moduli space. 
In particular, the pre-potential \eqref{orientifold_pre} contains 
contributions from two rather extreme regimes: the large volume corrections in 
$R$, and the logarithmic corrections in $v^i,b^i$ due to the singularity. However,
we cannot make $R$ extremely large, nor $v^i,b^i$ extremely small as we have already
discussed in section \ref{axion_decay}.
Let us note that for a given Calabi-Yau geometry the $\cN=2$ pre-potential 
can be computed much more explicitly than \eqref{orientifold_pre}. 
However, also additional perturbative $\cN=1$ 
corrections might alter the precise form of the effective theory. 
These perturbative corrections 
are believed to not depend on the axions. Hence, in case they do not
cancel the $\cN=2$ effects, an effective theory with light axions 
might still be accessible.\footnote{I like to thank S.~Kachru for discussions on this point.}
 Therefore, even though the following analysis
appears rather explicit, its results should be  interpreted with care and further study will be required
to gain a solid picture.

For a given pre-potential the $\cN=1$ K\"ahler coordinates are computed using \eqref{def-T}. 
Clearly, the complex dilaton $\tau$ and the coordinates  
\beq \label{G^i_coni}
   G^i = c^i +i e^{-\phi} \R\, t^i_- = c^i - i e^{-\phi} b^i\ ,
\eeq
do not depend on the form of the pre-potential. In our set-up,
we denote by $T_{\rm R}$ the coordinates corresponding to the larger cycles of radius 
$R$. Here R is not an index, but we will later work in a toy model with several $T_{\rm R}$
of the same size and include appropriate factors to label this degeneracy.
The definition of $T_{\rm R}$ depends on the pre-potential \eqref{orientifold_pre} in a 
rather simple way, since we only kept the classical terms in $\cF$. 
However, the K\"ahler coordinates $T_i$ associated with $v^i$ contain
the terms $\R(\partial \cF/ \partial {t^i_+})$ 
and hence will receive $\alpha'$ corrections form the logarithmic corrections.
Using the pre-potential \eqref{orientifold_pre} in \eqref{def-T} we derive
\bea \label{def-T_R}
   T_{\rm R} &=& - \rho_{\rm R} + i e^{-\phi} \big(\tfrac{1}{2} (R/\ell_s)^4 - \sum_j \big[(v^j)^2 - (b^j)^2 \big] \big)\ ,\\
   T_i &=& - \rho_i + i e^{-\phi} \big(-2(R/\ell_s)^2 v^i +v^i+\big[ -2 b^i  \arg(t^i) + 2 v^i \log |t^i| \big]\big)\ , \nn 
\eea
where we abbreviated $t^i  = -b^i + iv^i$.
Finally, the K\"ahler potential $K_{\rm q}$ can be evaluated using \eqref{orientifold_pre}.
This is straightforward, since the quantum corrected volume \eqref{toy_V_N} only depends on $v^i$
and the absolute value of $t^i$. The B-field $b^i$ appears only quadratic and 
the sign flip of $b^i$ in the last two terms of \eqref{orientifold_pre} only yields a factor of $2$.  
The expression for $K_{\rm q}$ is simply
\beq \label{K_qconif}
  K_{\rm q} = - 2 \log \Big[e^{-2 \phi} \big(  \tfrac{4}{3}  (R/\ell_s)^6  -8 (R/\ell_s)^2  \sum_i (v^i)^2 + 4  \sum_i \big( |t^i|^2 \log |t^i| +(v^i)^2 \big)\big) \Big]\ .
\eeq
The combination in the brackets is precisely $e^{-2\phi} \cV$, where $\cV$ is the quantum volume of $Y$, and 
should be compared to the earlier expression \eqref{toy_V_N} in section \ref{multiax}.
Expressing $K_{\rm q}$ as a function of the $\cN=1$ K\"ahler coordinates $\tau, G^i, T_i$ and $T_{\rm R}$
is considerably harder. In order to do that we would have to solve $\tau=C_0+ie^{-\phi}$,
as well as eqns.~\eqref{G^i_coni} and \eqref{def-T_R}
for $e^{-\phi},R,v^i,b^i$ and insert the result into \eqref{K_qconif}.
Fortunately, we will not need an the explicit expression for $K_{\rm q}$
as a function of the $\cN=1$ coordinates.

At least in principle, we are now in the position to compute the 
$\cN=1$ scalar potential \eqref{scalarpot} induced by a 
superpotential due to background fluxes and non-perturbative 
effects. A  non-perturbative superpotential of the form
\eqref{D3_superpotential}, \eqref{D3_superpotential_expand} can potentially 
stabilize the moduli $T_{\rm R},T_i$ and $G^i$. The flux superpotential depends on the 
complex dilaton $\tau$ and the complex structure moduli.
Altogether, the superpotential is a function of all bulk moduli of the $\cN=1$
effective theory and one expects that the scalar potential will admit 
minima  in which all fields can settle to their vacuum values. First 
the heavier fields will roll quickly down to their minima, while only 
later the light degrees of freedom will follow.  If the axions are indeed 
the lightest fields and the other fields are fixed to the desired 
values, the effective theory could allow the $N$-flation scenario of section \ref{rev_N} .

Given the $\cN=1$ characteristic data it is still not straightforward 
to explicitly check if our set-up admits an epoch of axion inflation. 
The reason for this is of technical nature. Firstly, the general 
expressions for the derivatives of the K\"ahler potential summarized 
in appendix~\ref{N=1_recall} are complicated functions of the pre-potential
\eqref{orientifold_pre}. In this work,
we will simplify the computations and evaluate the potential using 
the leading results \eqref{decay} for the axion decay constants 
and hence the K\"ahler metric.\footnote{For a small number of fields 
it can be checked numerically that this is a reasonable approximation. It 
becomes more accurate with increasing volume $\cV$.}
Secondly, since the fields $\tau,G^i,T_i$ and $T_{\rm R}$
mix both in the K\"ahler potential as well as in the superpotential it 
is hard to determine an effective theory for any subset of fields.
In order to nevertheless proceed, we will have to assume, similar to 
the approach of KKLT \cite{KKLT}, that we can stabilize
the fields in two steps. In a first step, the dilaton and complex structure moduli 
are stabilized using background fluxes. Later on,
within the effective theory for the fields $T_{\rm R},T_i$ and $G^i$, we will 
address the stabilization of the remaining bulk fields 
in an up-lifted non-supersymmetric vacuum. One expects that this will 
give at least a qualitative picture of the $N$ conifold toy model.

Let us begin by utilizing the flux superpotential to 
fix the complex structure moduli  $z^k$ and the dilaton $\tau$
such that
\beq \label{compl_struc}
   D_{z^k} W = 0 \ , \qquad \qquad D_{\tau} W =0\ .
\eeq
As already mentioned above, our set-up is so 
complicated that the second condition in \eqref{compl_struc}
will receive corrections through the derivative of the K\"ahler potential 
\eqref{K_tau}. However, we will assume that 
we can choose the fluxes such that these corrections are subleading 
and we can fix $\tau$ nevertheless to an appropriate value. 
The effective potential for the remaining fields is obtained by 
inserting \eqref{compl_struc} into the general expression \eqref{scalarpot}.
Setting $S^A = (G^i,T_{\rm R},T_i)$ one finds
\beq \label{V_effective}
   V_{\rm eff} =  e^K(K^{S^A \bar S^B} D_{S^A} W \overline{D_{S^B} W} - 3|W|^2) + V_{\rm up}\ .
\eeq
As in the scenarios of refs.~\cite{KKLT,KKLMMT, BKQ}, we will add an up-lift term 
of the form $V_{\rm up}={\kappa}/{\cV^\alpha}$ to $V(S)$, where $\kappa$ is a tunable 
constants and $\alpha$ is a rational number of order unity depending on the 
source of the up-lifting energy. By fine-tuning $\kappa$, the positive contribution $V_{\rm up}$ 
can be utilized to lift a vacuum at negative vacuum energy to a very small positive 
cosmological constant. This up-lifting does typically not significantly change 
the location of the vacua  and the local shape of the potential close to the lifted 
minimum. Therefore, we will set the up-lifting term in \eqref{V_effective} to be zero in 
order to simplify our analysis. Clearly, in order to make contact to section \ref{rev_N}
it eventually has to be included  and appropriately fine-tuned such that the cosmological
constant $C$ in \eqref{sum-pot} is negligible during inflation.

Following ref.~\cite{KKLT} and using \eqref{D3_superpotential}, \eqref{D3_superpotential_expand}, the 
effective instanton superpotential in \eqref{V_effective} takes the form 
\beq \label{reduced_super}
  W = W_0 + \sum_{j=1}^N e^{i T_j} + e^{iT_{\rm R}} + e^{-1/g_s} \sum_{j=1}^N e^{-iG^j} e^{i T_{\rm R}}\ ,
\eeq
where $W_0$ is a constant depending on the values of the background fluxes.
We observe that only the last term in \eqref{reduced_super} 
depends on the R-R two-form axions. At small string coupling and 
$\I T_{\rm R}>1$ this term is strongly suppressed in comparison 
to the other contributions in $W$. Since the masses of the fields 
are weighted by the instanton action appearing in the superpotential,  
this simple observation leads us to expect that the masses of the fields 
in $G^i$ might be lower than the masses of $T_{\rm R},T_i$. Moreover, in 
the absence of the $G^i$ dependent term in $W$, the axions $\R\, G^i=c^i$ 
are actually massless. However, due to the $b^i$ dependent 
corrections to the K\"ahler potential the fields $\I\, G^i=-b^i/g_s$ can acquire a
mass even if they do not appear in the superpotential. 
As we will see below, they can be already stabilize to 
their vacuum in the absence of the $G^i$ dependent term in $W$.

In the following we numerically 
investigate our simple toy model in more detail. Our goal is to show that 
the included string world-sheet corrections alter the vacuum structure 
of the theory and can result in a theory of light axions. This should be
viewed as a qualitative result, since the precise values presented in 
the following are only valid with our assumptions and an 
appropriate fine-tuning.
For our numerical example, we will tune the fluxes such that 
 $g_s = 0.16$ and $W_0 = -0.12$.
Moreover, we will consider a set-up of $100$ larger cycles 
of identical radius $R$. To each of these cycles we assume that
there are coupling $8$ different resolved conifold singularities.
The orientifold projection should identify these pairwise, such 
that we have $400$ conifold pairs and $N=400$ associated 
axions from the R-R two-form.
 In order to get a qualitative picture of the resulting potential we will restrict to a model with 
effectively four real fields: the radius $R$ of the larger cycles and the three fields associated 
to each conifold pair
\beq
   v\approx v^i\ ,\qquad \qquad b\approx b^i\ , \qquad \qquad c \approx c^i\ ,
\eeq
where $v,\ b$ parametrize the resolving volumes 
and B-fields and $c$ are the R-R two-form axions.\footnote{It is straightforward 
to also include the R-R four-form axions, but we will omit them for simplicity.}
As in section \ref{multiax}, 
it is important to also carefully keep track of factors labeling the degeneracy of the large 
cycles and the associated conifold pairs.

In the computation of the effective potential \eqref{V_effective} we use the 
leading results for the K\"ahler metric \eqref{decay}, the superpotential \eqref{reduced_super} 
as well as the derivatives of the K\"ahler potential \eqref{K_tau}.
Explicitly, one computes
\bea
  K_{T_{\rm R}} &=& 4i g_s\, { (R/\ell_s)^2} \, \cV^{-1}\ ,\qquad \qquad K_{T_{j}} = 4i g_s\, { v} \, \cV^{-1} \ , \\
  K_{G^j} &=& -4 i g_s\, \big(-2(R/\ell_s)^2 b +b+\big[ 2 v  \arg(-b+iv) +  b \log (b^2+v^2) \big] \big)\, \cV^{-1}\ . \nn
\eea
In a next step the effective potential $V(R,v,b,c)$ can be minimized numerically. 
This yields an anti-de~Sitter  minimum at 
$\langle R \rangle= 1.208$, $\langle v \rangle =-0.1225$, $\langle b \rangle =0.2564$ and $\langle c \rangle =0$, which we have plotted 
in appendix \ref{Plots}, Figures \ref{bR_plot}, \ref{vb_plot} and \ref{vR_plot}.
Inserting these vacuum values for $R,v,b$ and $c$ into 
the K\"ahler coordinates \eqref{def-T_R} and \eqref{G^i_coni}, 
this minimum corresponds to \footnote{Recall that we have $100$ coordinates $T_{\rm R}$ with each $4$
associated conifold pairs. This implies that the sums in the definition \eqref{def-T_R} of each 
$T_{\rm R}$ runs from $j=1\ldots 4$.}
\beq \label{min_coord}
 \langle \I\, T_{\rm R} \rangle =7.924\ , \qquad \langle \I\, T_j \rangle =  12.039\ ,\qquad \langle G^j \rangle = 0 - 1.603 i \ .
\eeq
We observe that these vacuum values are in the a field range consistent with the assumption 
that D-brane effects are subleading in the K\"ahler potential. The respective D1 and D3
instanton actions are sufficiently suppressed by  $-\I\, G^j$ and $\I\, T_{\rm R},\ \I\, T_j$.
It is also straightforward to compute the string-frame volume in \eqref{K_qconif} at this minimum,
$\cV =205.6$. This is not extremely small which implies that also the axion decay constant 
of each R-R two-form axion $c$ is not expected to be extremely close to the Planck scale.
Explicitly, the axion decay constants are computed by inserting the vacuum values \eqref{min_coord} 
into equation \eqref{decay} for $C_2$ yielding
\beq \label{fC2}
   f_{C_2} = 0.02\ M_P\ .
\eeq
Clearly, this is too small to match the cosmological data. However, $f_{C_2}$ can 
made larger by fine-tuning the value of $g_s,\ W_0$ and considering a more 
sophisticated geometry. For the following qualitative analysis the precise value 
of $f_{C_2}$ will not be directly relevant.

\begin{figure}[h!]
\leavevmode
\begin{center}
\begin{minipage}[b]{7.3cm}
\includegraphics[height=6.0cm]{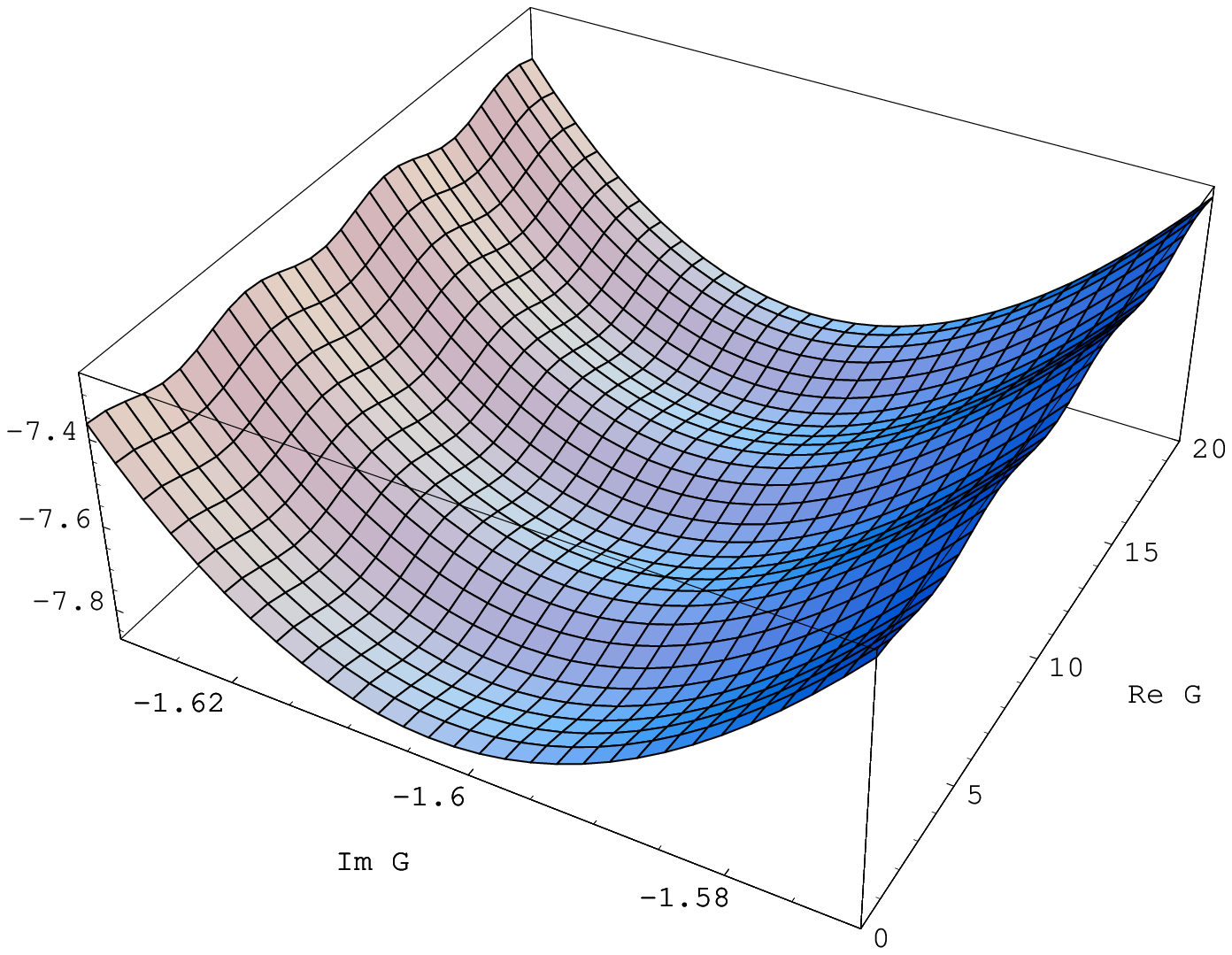}
{\caption{\small Axion valley potential $V_{\rm eff}(G)$ (multiplied by $10^{11}$) 
for axion $\R\, G$ and its non-axionic partner $\I\, G$, at fixed $\langle T_{\rm R}\rangle$
and~$\langle T_{i} \rangle$. \hspace*{5cm} \label{AxionV_with}}}
\end{minipage}
\qquad
\begin{minipage}[b]{7.3cm}
 \includegraphics[height=6.0cm]{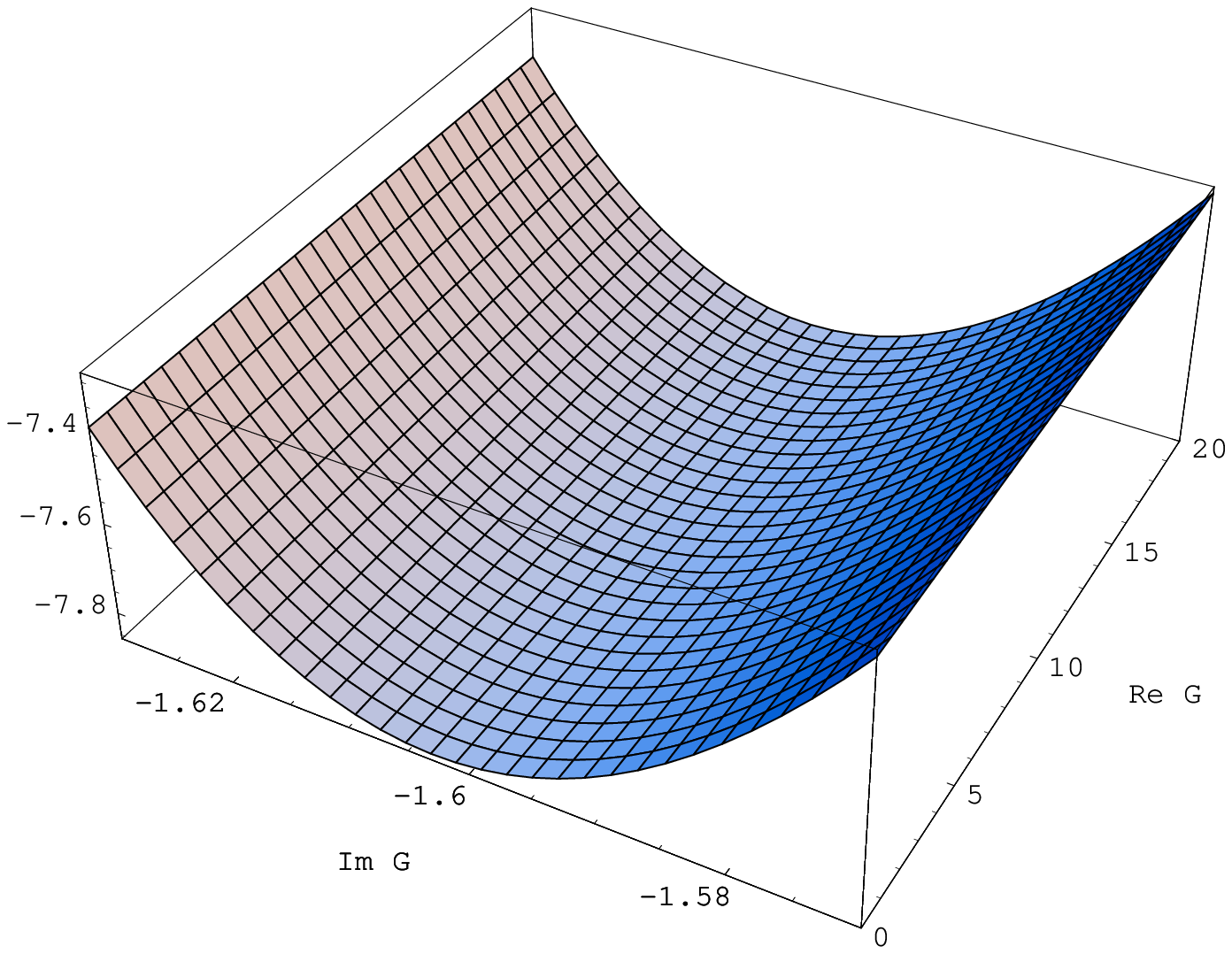} 
 {\caption{\small Axion valley potential $V_{\rm eff}(G)$ in the 
 absence of a superpotential depending on $G$. The axion direction 
 $\R\, G$ is flat, while $\I\, G$ still has a minimum.\label{AxionV_without}}}
 \end{minipage}
\end{center}
\vspace*{-.5cm}

\end{figure}

Let us now focus on the coordinates $G^j$ containing 
the R-R two-form axions $c$.
In order to get a picture of the axion potential, we first 
have to express $V_{\rm eff}(b,c,v,R)$ as a function of the $\cN=1$
K\"ahler coordinates $T_i,T_{\rm R}$ and $G^i$ such that $V_{\rm eff}(T_i,T_{\rm R},G^j)$. 
This can be only done numerically, since the coordinate definition \eqref{def-T_R} and \eqref{G^i_coni}
are highly non-linear.
In a next step we fix $T_R$ and $T_i$ to their vacuum values  \eqref{min_coord}
and plot the scalar potential as a function of the axion $\R\, G^i = c$
and its non-axionic partner $\I \, G^i = -b/g_s$. The
result is shown in Figure~\ref{AxionV_with}. The potential takes the 
form of a valley. It is steep in the direction of the non-axionic field $b$,
but shallow in the direction of the axion $c$.  This is very similar to the 
form of the axion valley used by Kallosh in ref.~\cite{Kallosh} to model natural inflation 
in supergravity. In order to check that there is indeed a mass hierarchy between the 
axions $\R\, G^i$ and their non-axionic partners $\I\, G^i$, we can switch of 
the superpotential corrections depending on $G^i$ by setting  
the last term in \eqref{reduced_super} to zero. The potential $V_{\rm eff}(G)$
at the vacuum values \eqref{min_coord} for  $T_R$ and $T_i$ 
is plotted in Figure~\ref{AxionV_without}.
We conclude that the minimum stabilization of $\I\, G$ arises though 
corrections in the K\"ahler potential.
The mass hierarchy between the axion and its non-axionic partner is 
crucial for inflation.  It ensures that first the non-axionic field 
settles to its minimum leaving an effective theory for the light axion
discussed in section \ref{rev_N}. Note that this hierarchy is supported by
the world-sheet corrections to the quantum volume $\cV$.

Let us end this section by computing the axion potential near the minimum 
\eqref{min_coord}.
For the canonically normalized R-R two-form axions 
$\theta^i = f_{C_2} c^i$ one finds 
\beq \label{axion_eff_cone}
   V_{\rm eff}(\theta^i)/M_P^4 = -7.86\cdot 10^{-11}  + 
    \sum_{i=1}^N 5.07 \cdot 10^{-16}\big(1- \cos( 50\, \theta^i/M_P)\big) \ ,
 \eeq
where we have taken all $N=400$ axions to be of similar size $\theta \approx \theta^i$.
The effective potential \eqref{axion_eff_cone} is precisely of the form \eqref{sum-pot}. 
However, note that our model was not sufficiently fine-tuned to obtain 
a pseudo-realistic cosmology. Firstly,
the cosmological constant in \eqref{axion_eff_cone} is still negative and has to be
lifted to a small positive value by including the up-lifting term $V_{\rm up}$
in the effective potential \eqref{V_effective}. As in ref.~\cite{KKLT}, this additional 
term is independent of all R-R axions and will not change the location of the 
minimum significantly. Thus our discussion of the mass hierarchy still remains 
valid in the up-lifted vacuum. Secondly, the axion decay constants \eqref{fC2} 
are still too small and we have to further improve our set-up to find the 
desired cosmology. In any case, as already stressed above, the given form 
of the effective theory should only be interpreted qualitatively, 
since we worked with the simple pre-potential \eqref{orientifold_pre}
and applied several approximations in deriving $V_{\rm eff}$. Our analysis 
indicates that there might exist compactifications with a large number
of axions as the lightest bulk moduli. A more intensive study of explicit 
models will reveal whether these scenarios are indeed 
suitable to obtain axion $N$-flation.

\section{Conclusions}

In this paper we discussed the possibility of a string theoretical 
embedding of cosmological inflation driven by a large number of axions. 
Such scenarios use the fact that in the dimensional reduction 
of string theory  to four space-time dimensions a vast 
number of scalar fields arise in the low energy theory. In particular,
we considered type IIB string theory on a compact Calabi-Yau 
manifold with many non-trivial two-cycles. To each of these 
two-cycles an axion from the R-R two-form and an axion 
from the R-R four-form can be associated. The 
effective theory for the axionic fields will sensitively depend 
on the values of the geometric moduli, i.e.~the volume of the
two-cycles, as well as the NS-NS B-field. We have argued that a 
possible realization of axion inflation might only exist in special 
corners in the landscape of vacua. In these regimes various stringy
effects become relevant and have to be included. In this work, we 
have made first steps one region in the $\cN=1$ landscape, where 
axion inflation might be realized.

In recent years, $\cN=1$ type IIB compactifications with  
all volumes stabilized at scales much larger than string scale have been 
investigated intensively \cite{review_flux}.
For vacua in this regime of the field space, stringy corrections, such
as wrapping stings and D-branes, play a sub-leading role in the derivation of the 
kinetic terms of the axions. Non-perturbative D-brane effects can induce
a potential exponentially suppressed by the large volume 
of the cycles times the inverse of a small string coupling. We have argued 
that such scenarios are not suitable for axion inflation. They naturally 
admit very small axion decay constants which become even smaller 
if the number of axions increases. Moreover, due to the strong 
exponential suppression, the scale of inflation is typically too low for 
semi-realistic scenarios. This conclusion lead us to the consideration 
of manifolds with small or vanishing cycles.

In compactifications with cycles smaller than string scale the 
caveats of the large volume compactifications can be avoided. 
To obtain such geometries, we considered 
resolutions of singularities supported by a volume and 
a NS-NS B-field. The standard example of such a blown-up 
singularity is the resolved conifold.
In the vicinity of small cycles, new corrections will become 
relevant and alter the effective theory. We have discussed
a subset of such correction in $\cN=2$ compactifications of 
type IIB string theory on a Calabi-Yau manifold. Including the 
leading singular non-perturbative string world-sheet contributions,  
we have argued that the axion decay constants can take values 
close to the Planck scale also for scenarios with many axions. 
In $\cN=1$ compactifications further perturbative and non-perturbative corrections can
become relevant and might alter the structure of the effective four-dimensional 
theory. However, in case these do not cancel the $\cN=2$ effects, large 
axion decay constants will remain accessible also in these less supersymmetric 
scenarios. 

In addition to being close to the Planck scale, $N$-flation also 
requires the axion decay constants to be independent of the axions 
themselves. Corrections depending on the R-R axions are believed to only 
arise from non-perturbative D-brane effects. If the small cycles 
remain to be of finite size and we work at sufficiently small string coupling, 
D-brane corrections are subleading in the axion decay constants, since 
the instanton action is larger by a factor of the inverse string coupling.
This will also be the case for vacua in an $\cN=1$ supergravity theory. 
In a consistent analysis, the vacuum values of the $\cN=1$ moduli fields 
have to be inside an appropriate field range to ensure that D-brane instantons 
are subleading in the K\"ahler potential encoding the kinetic terms of the scalar fields.

One of the remaining complications in realizing $N$-flation 
is to ensure that there indeed exists an effective theory for axions
with appropriate masses during inflation. These masses have
to be lower than the masses of their non-axionic partner and other 
bulk moduli fields, but still sufficiently large to match the observed 
density perturbations during inflation. Again, this forces us to work away from the large 
volume regime, where at least some of the moduli masses
are suppressed by large instanton actions. A careful investigation 
of the effective potential for the moduli is necessary to study 
the vacua of the theory. In this 
work we briefly discussed effective $\cN=1$ potentials 
arising from D-instantons and gaugino condensates 
on space-time filling branes. We pointed out that for 
potentials induced by large rank gaugino condensates on D5 branes,
the quadratic region of the axion potential can stretch over the 
entire accessible field range. In practice, D1 corrections 
arising through the pre-factors of the D3-instantons are 
of particular interest. If these are the leading axion-dependent
contributions to the $\cN=1$ superpotential, a mass hierarchy 
can be ensured due to the suppression by both the D1 and D3 instanton 
action. 

To investigate the properties of the 
effective theory more explicitly, we discussed the
embedding of axion inflation into $\cN=1$ Calabi-Yau 
orientifold compactifications with O3 and O7 planes. 
We showed that the $\cN=1$ characteristic data remain 
calculable even in the case that the non-perturbative $\cN=2$
string world-sheet corrections are included. Utilizing 
a flux superpotential together with a superpotential 
from non-perturbative D-brane effects a potential for all bulk 
moduli fields is generated. We illustrated that a theory of 
light axions could exist, if the axion dependence is 
suppressed in the superpotential and the non-axionic
partners of the axions in the $\cN=1$ chiral multiplet is 
stabilized due to the string world-sheet corrections 
to the K\"ahler potential. 
In an optimistic scenario, 
one can hope that such an effective theory of a large number 
of relatively light axions from the R-R forms will survive 
also further perturbative corrections. 

Even though an explicit embedding of $N$-flation into 
string theory still remains to be constructed, the scenarios
outlined and studied in this work might provide a promising 
route to achieve this goal. Likely, such an embedding will not solve
intrinsic issues related to the fine-tuning of initial conditions in 
chaotic and natural inflation with many inflatons. 
However, it might provide a way to 
accommodate possible future observations of primordial gravitational 
waves in a string theoretic model.

\section*{Acknowledgments}

I am particularly in debt to S.~Kachru, R.~Kallosh and A.~Klemm for many enlightening discussions,
continuous  correspondence and useful comments on the draft. For very useful discussions
and comments on the draft, I would like to thank G.~Shiu and B.~Underwood.
I am also grateful to M.~Aganagic, I.~Ellwood, O.~Ganor, G.~Geshnizjani, J.~Heckman, H.-P.~Nilles, E.~Pajer,
F.~Saueressig, S.~Shandera and J.~Walcher for communicating their insights on related
topics.  This work was partially supported by the
European Union 6th framework program MRTN-CT-2004-503069
``Quest for unification", MRTN-CT-2004-005104 ``ForcesUniverse",
MRTN-CT-2006-035863 ``UniverseNet" and 
SFB-Transregio 33 ``The Dark Universe" by the DFG, the NSF CAREER Award No. PHY-0348093,
DOE grant DE-FG-02-95ER40896, a Research Innovation
Award and a Cottrell Scholar Award from Research
Corporation.

\newpage

\appendix

\noindent {\bf \LARGE Appendices}

\section{Axions in $\cN=2$ Calabi-Yau compactifications \label{gen_axion_const}}

In this appendix we recall the effective action of type IIB string 
theory compactified on a Calabi-Yau manifold $Y$. Our analysis 
will not necessarily take place in the large volume limit, such that
stringy corrections have to be taken into account. 
Clearly, we will not be able to incorporate all of these corrections 
into our analysis. However, for small string coupling $g_s$ there is a regime
in parameter space in which we have control over the effective 
theory as discussed in sections \ref{res_cone} and \ref{multiax}. In particular, contributions
due to D-branes in the type IIB theory are subleading for small $g_s$.
Nevertheless, the four-dimensional 
physics can be corrected by contributions from string world-sheets.
In an $\cN=2$ compactification such $\alpha'$ corrections are
encoded by a holomorphic  pre-potential $\cF$
which can be expanded around the desired point 
in the moduli space. In general, the pre-potential $\cF$ 
is a function of the complexified K\"ahler structure deformations
$t^A$ defined in \eqref{def-t}. We also introduce homogenous coordinates $X^{\hat A} = (X^0,X^A)$
and write
\beq
  F(X^{\hat A}) \equiv (X^0)^2 \cF(t^A) \ ,\qquad \qquad t^A = X^A/X^0\ .
\eeq
All four-dimensional $\cN=2$ data will be given as a function of the 
homogenous pre-potential $F(X)$ and hence as a function of $\cF(t)$.

Recall that compactifying type II string theory on a Calabi-Yau manifolds
leads to an $\cN=2$ supergravity theory with $h^{(1,1)}+1$ hypermultiplets.
The scalars in these hypermultiplets are the complex scalars $t^A$ given in \eqref{def-t},
the lowest modes of the ten-dimensional dilaton and type IIB R-R forms $C_0,\ C_2,\ C_4$.
More explicitly, this includes the universal hypermultiplet $(\phi,C_0, \rho_0, c^0)$, where 
$\rho_0,\ c^0$ are the duals of the four-dimensional two-form part in $B_2,C_2$.
In addition there are $h^{(1,1)}$ hypermultiplets $(t^A, \rho_A, c^A)$ with
\beq
   c^A = \tfrac{1}{2 \pi} \int_{\cC_A} C_2 - C_0 B_2\ ,\qquad \quad 
   \rho_A = \tfrac{1}{2 \pi} \int_{\tilde C^A} C_4 - B_2\wedge C_2 +\tfrac{1}{2} C_0 B_2 \wedge B_2\ ,
\eeq
where $\cC_A$ and $\tilde \cC^A$ are harmonic two- and four-cycles of $Y$. Note
that we are slightly abusing the notation compared to \eqref{C-expansion}, since $c^A,\rho_A$ now also include
corrections due to lower R-R forms. We combine $(\rho_0,c^0)$ and $(\rho_A,c^A)$ by writing 
$(\rho_{\hat A},c^{\hat A})$ with $\hat A=0 \ldots h^{(1,1)}$.
Having identified the $\cN=2$ hypermultiplets we  
turn to their effective action and moduli space metric.
From a Kaluza-Klein reduction one derives the effective Lagrangian \cite{FS}
\bea \label{q-metr}
\cL^{(4)} &=& (\partial D)^2 + G_{A \bar B}\, \partial t^A \partial\bar t^B
                               +\tfrac{1}{4}e^{4D}\big(dC_0 -(\rho_\Ah \partial c^\Ah - c^\Ah d\rho_\Ah) \big)^2 - \tfrac{1}{2} e^{2D} \text{Im} \cM_{\Ah \Bh} \partial c^\Ah \partial c^\Bh \nn \\
                              &&  - \tfrac{1}{2} e^{2D} (\text{Im} \cM)^{-1\, \Ah \Bh}
                                   \big(\partial \rho_\Ah - \R \cM_{\Ah \Ch} \partial c^\Ch \big)
                                   \big(\partial\rho_\Bh - \R
                                   \cM_{\Bh \Dh} \partial c^\Dh \big)\ ,
\eea
where $G_{A \bar B} = \partial_{t^A} \partial_{\bar t^B}K$ is the metric on the space of complexified 
K\"ahler structure deformations $t^A$ and given in terms of the K\"ahler potential \eqref{general_K}. 
The complex coupling matrix $\cM_{\Kh \Lh}$ 
appearing in \eqref{q-metr} depends on $t^A, \bar t^A$
and is defined as
\beq \label{def-cM}
   \cM_{\hat A \hat B} = \overline{F}_{\Ah \Bh}+2i \frac{(\text{Im} F_{\Ah \Ch}) X^\Ch
   (\text{Im} F_{\Bh \Dh})X^\Dh}{X^{\hat E} (\text{Im} F_{\hat E \hat F} )\ ,
    X^{\hat F}}\ ,
\eeq
where $F_{\Ah \Bh}=\partial_{X^\Ah} \partial_{X^\Bh} F$.
Finally, the Lagrangian \eqref{q-metr} contains the four-dimensional 
dilaton $D$ defined in terms of the ten-dimensional dilaton $\phi$ according to
\beq \label{4d-dilaton}
   e^{D} = e^{\phi} \cV^{-\frac{1}{2}}\ ,
\eeq
where $\cV(t,\bar t)$ is given in \eqref{general_K}.

Using the Lagrangian \eqref{q-metr} it is straightforward to read off the 
axion decay constants for the axions $b^A$, $c^A$ and $\rho_A$. 
For the NS-NS B-field axions $b^A$ we find 
\beq
   B_2: \qquad  \frac{f_{AB}^2}{M_P^2} = \frac{1}{\pi} G_{A \bar B}\ ,  
\eeq
just as given in \eqref{decay}. The expression for the R-R two- and four-form axions $c^A, \rho_A$ 
are more complicated 
\bea \label{general_decay}
   C_2: \qquad  \frac{f_{AB}^2}{M_P^2}  &=& - \frac{g_s^{2}}{2\pi \cV} \left(\text{Im} \cM_{A B} 
     + \R \cM_{A \Ch}  (\text{Im} \cM)^{-1\, \Ch \Dh} \R \cM_{\Dh B}  \right)\ ,\\
     C_4 : \qquad \frac{f_{AB}^2}{M_P^2}  &=& - \frac{g_s^{2}}{2\pi \cV} (\text{Im} \cM)^{-1\, A B} \ .
\eea
These expressions appear to be different from the ones used in the main text \eqref{decay}.
However, due to the underlying special geometry we can use
\beq
  - 2 \cV^{-1} \text{Im} \cM_{A B}  = G_{A \bar B} + \ldots \ .
\eeq
In other words, the metric $G_{A\bar B}$ is the leading contribution to these axion decay constants
and \eqref{general_decay} reduces to \eqref{decay}. Of course, the study 
of the axion decay constants performed in the main text can be repeated with the 
general expressions \eqref{general_decay}.

\section{The $\cN=1$ K\"ahler metric \label{N=1_recall}}

In this appendix we summarize some useful formulas
allowing to derive the derivatives of the K\"ahler 
potential $K_{\rm q}$ given in \eqref{Kqgeneral}. Our summary  
will follow ref.~\cite{GL2}.
Let us recall, for completeness, that $K_{\rm q}$ takes the 
form 
\beq \label{recall_Kq}
   K_{\rm q}(\tau,G,T) = - 2 \ln \big[i e^{-2\phi} \big( 2(\cF -\bar \cF)-(\cF_A +\bar \cF_A)(t^A-\bar t^A) \big) \big]\ ,
\eeq
with complex coordinates 
\beq \label{recall_N=1coords}
   \tau = C_0 + i e^{-\phi}\ ,\qquad \quad    G^a = c^a + ie^{-\phi} \R\, t^a \ ,\qquad \quad 
   T_\alpha =-\rho_\alpha + i e^{-\phi} \R\, \cF_{\alpha} \ .
\eeq 
In the equations \eqref{recall_Kq} and \eqref{recall_N=1coords}, we have denoted by $\cF_A=(\cF_\alpha,\cF_a)$
the derivatives of a general pre-potential $\cF$ with respect to 
the special coordinates $t^\alpha,t^a$. Note that $t^\alpha$ and $t^a$ are associated 
to the positive and negative two-cycles in the eigenspace of the orientifold projection \eqref{H^2split}.

With our conventions, the first derivatives of the K\"ahler potential \eqref{recall_Kq} 
are given by  
\bea \label{K_tau} 
   K_{\tau} &=&-4i \, e^{-\phi} \I (2\cF - t^A \cF_A)\ e^{K_{\rm q}/2}\ ,\qquad \quad K_{G^i} =-4i \,  e^{-\phi} \I\, \cF_a\ e^{K_{\rm q}/2} \ , \nn \\
   &&\hspace{3cm} K_{T_\alpha} = 4i\,  e^{-\phi} \I\, t^\alpha \ e^{K_{\rm q}/2}\ .
\eea
We note that the K\"ahler coordinates \eqref{recall_N=1coords} are functions of the real parts $\R\, t^a, \R \cF_\alpha$
while the first derivatives of the K\"ahler potential are the imaginary part of $\cF$ and its derivatives.

The K\"ahler metric and its inverse can be also expressed as functions of a 
pre-potential $\cF$. Let us denote $N^{\hat a}=(\tau,G^a)$, where the complex
dilaton $\tau$ is identified with $N^0$.
One has
\bea
   K_{T_\alpha \bar T_\beta} &=& - 2 e^{2D} (\I \cM)^{-1\, \alpha \beta}\ ,\qquad \qquad 
   K_{T_\alpha \bar N^\ah} = 2 e^{2D}   (\I \cM)^{-1\, \alpha \beta} \R \cM_{\beta \ah}\ , \nn \\
   K_{N^{\hat a} \bar N^{\bh}} &=& -2 e^{2D} \big( \I \cM_{\ah \bh} +   \R \cM_{\ah \alpha} (\I \cM)^{-1\, \alpha \beta} \R \cM_{\beta \bh} \big)\ ,
\eea
with inverse 
\bea
   K^{T_\alpha \bar T_\beta} &=& - \tfrac{1}{2} e^{-2D} \big( \I \cM_{ \alpha \beta} +   \R \cM_{ \alpha \ah} (\I \cM)^{-1\, \ah \bh} \R \cM_{\bh \beta}\big)\ , \\ 
    K^{N^\ah \bar N^\bh} &=&- \tfrac12 e^{-2D} (\I \cM)^{-1\, \ah \bh}\ ,\qquad \qquad  
    K_{T_\alpha \bar N^\ah} =- \tfrac12 e^{-2D}   (\I \cM)^{-1\, \ah \bh} \R \cM_{\bh \alpha}\ . \nn
\eea
In these expressions $\cM_{\hat A \hat B}$ is the complex matrix defined in \eqref{def-cM}
and $e^{D}$ is the four-dimensional dilaton given in \eqref{4d-dilaton}.

\section{Plots of minimum \label{Plots}}

This appendix contains the plots of the minimum for the toy model 
with $N=4$ conifold pairs discussed in section \ref{N_conifolds}.
The effective theory was evaluated using the pre-potential 
\eqref{orientifold_pre} at $g_s=0.16,\ W_0 = -0.12$. The numerically 
determined minimum is found at $\langle R \rangle= 1.208$, 
$\langle v \rangle =-0.1225$, $\langle b \rangle =0.2564$ and $\langle c \rangle =0$.
\begin{figure}[htp]
\leavevmode
\begin{center}
\includegraphics[height=6cm]{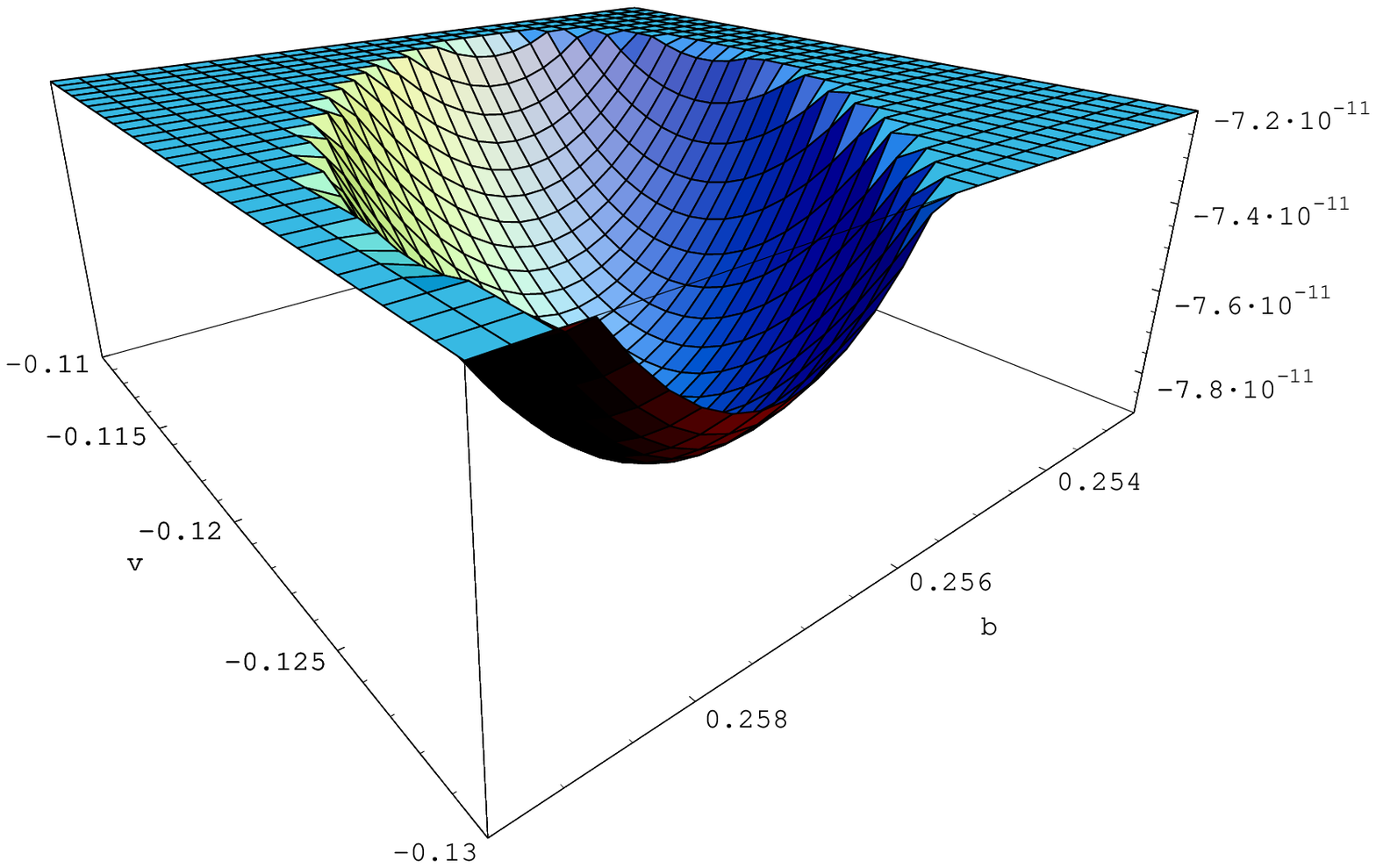} 
\end{center}
\vspace*{-.5cm}
\caption{\small Potential as function of conifold volume $v$ and  B-field $b$. $R,c$ are fixed to minimum. \label{vb_plot}}
\leavevmode
\begin{center}
\includegraphics[height=6cm]{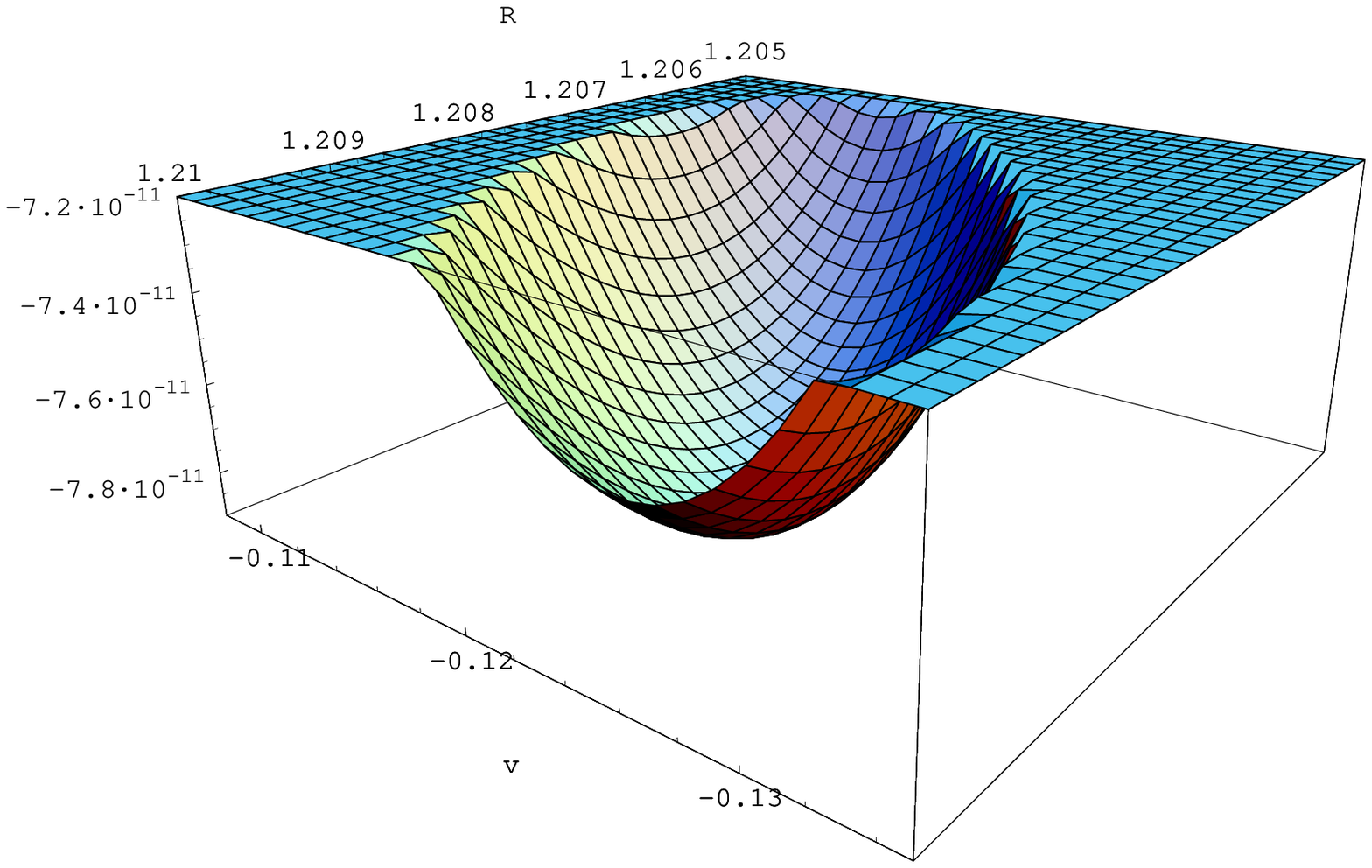} 
\end{center}
\vspace*{-.5cm}
\caption{\small Potential as function of radius $R$ and conifold volume $v$. $b,c$ are fixed to minimum. \label{vR_plot}
}
\leavevmode
\begin{center}
\includegraphics[height=6cm]{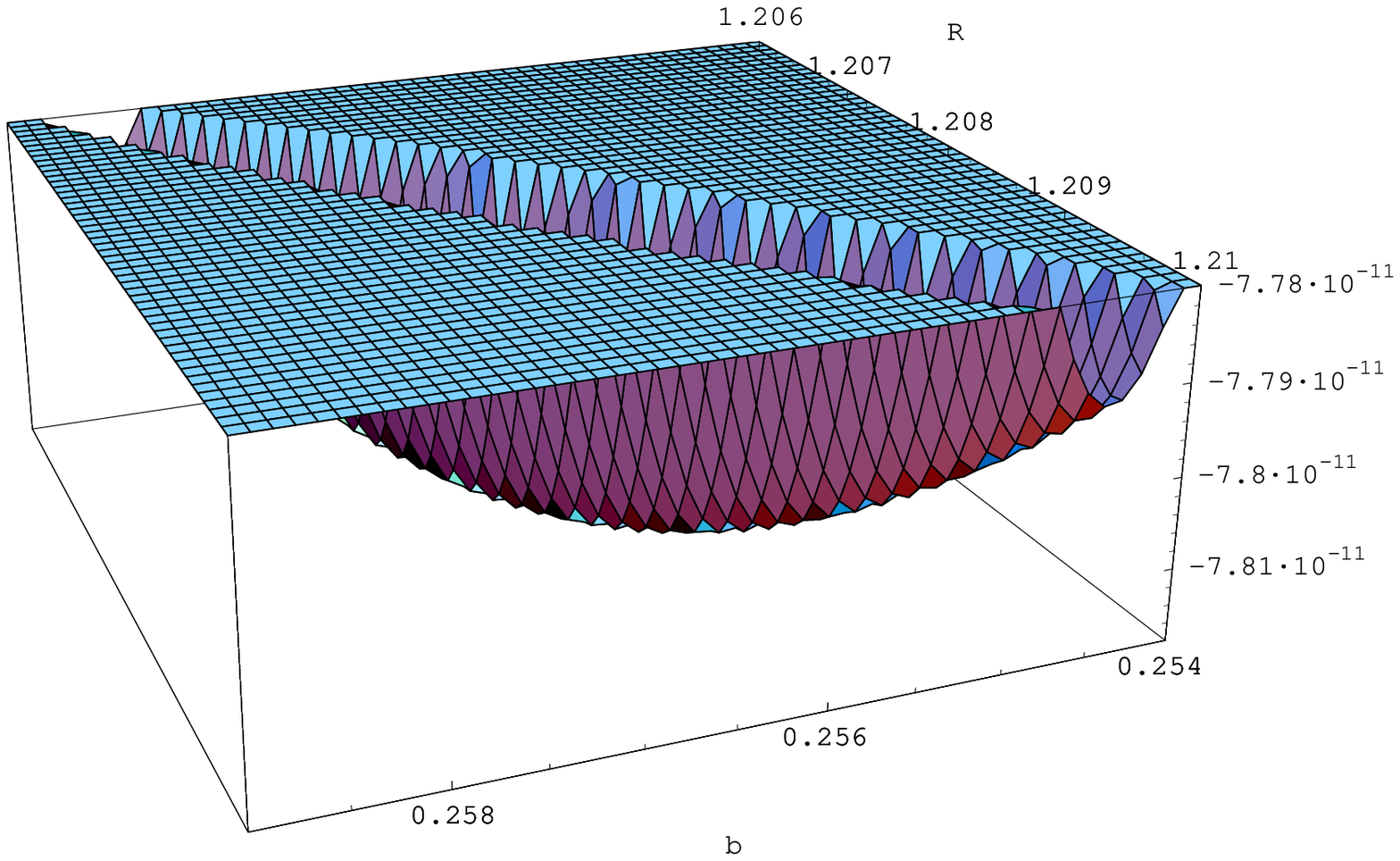}
\end{center}
\vspace*{-.5cm}
\caption{\small Potential as function of radius $R$ and conifold B-field $b$.  $v,c$ are fixed to minimum.\label{bR_plot} }

\end{figure}

\newpage
\newpage


\end{document}